%% file: main.tex
\documentclass[11 pt, letterpaper]{article}
\usepackage{setspace}
\usepackage{geometry}
\usepackage{graphicx}
\usepackage[flushleft]{threeparttable}
\usepackage{caption}
\usepackage{longtable}
\usepackage{mathtools, amssymb, nccmath}
\usepackage{amsmath}
\usepackage{bbm}
\usepackage{float}
\usepackage{setspace}
\usepackage[flushleft]{threeparttable}
\usepackage{natbib}
\usepackage[x11names]{xcolor}
\usepackage[
  colorlinks=true,
  citecolor=DodgerBlue4,   
  linkcolor=DodgerBlue4
]{hyperref}

\begin{document}

\title{How Election Shocks Impact Markets: Evidence from Sectoral Stock Prices\thanks{%
I am grateful to Serdar Birinci, Lawrence Christiano, Martin Eichenbaum, Jacob Gosselin, Diego Känzig, Jose Luis Lara, Mike McCracken, Joel Mokyr, Giorgio Primiceri, and Ramya Raghavan for helpful comments as well as participants at the Northwestern University 501 Seminar and Macro Ideation Session.}}
\date{\today }
\author{Aaron J. Amburgey\thanks{%
Northwestern University. Email: amburgey@u.northwestern.edu}}
\maketitle

\thispagestyle{empty}
\begin{abstract}
This paper examines the effects of U.S. presidential election cycles on sectoral stock markets. Using a high-frequency identification approach, I construct a novel ``election shock'' series, which captures exogenous surprises in election probabilities. Aside from election outcomes, the largest shocks are associated with events that are orthogonal to innovations in the macroeconomy, e.g., scandals and debates. These shocks have immediate effects on asset prices in sectors that are differentially impacted by the policy platforms of the two major U.S. political parties. In particular, shocks favoring Republican (Democratic) candidates increase (decrease) the asset prices in the energy and defense sectors, while decreasing (increasing) prices in the clean energy sector. These effects persist overtime.
\end{abstract}

\singlespacing

\newpage
\doublespacing

\setcounter{page}{1}
\setcounter{tocdepth}{0} 

\section{Introduction}
\label{sec1}

Increasing polarization in U.S. politics has sparked interest in the diverging economic implications of each of the major political parties’ policy platforms. For instance, recent Democratic presidential candidates have pushed for investment in the clean energy sector and a reduction in CO2 emissions, exemplified by policies like the Inflation Reduction Act (IRA) of 2022, while Republican candidates have supported the increased extraction of fossil fuels through expanded drilling and deregulation. Moreover, a large body of empirical literature demonstrates that news about changes in future fundamentals affects the economy today by altering expectations \citep[for a review, see][]{beaudry_news-driven_2014}. However, little work has been done to study the potential economic effects of news about future policy that is transmitted through election events.

In this paper, I investigate this mechanism by constructing a novel ``election shock'' series. To do so, I orthogonalize changes in daily election probabilities to financial and macroeconomic news. The resulting series captures exogenous surprises in the probability that each candidate is elected for the seven most recent U.S. presidential elections. Aside from election outcomes, the largest shocks are associated with notable recent election events, e.g., developments in the 2016 FBI investigation of Secretary Hillary Clinton's leaked emails and the 2024 debate between Donald Trump and Joseph Biden. The series' ability to capture variation associated with scandals and debates that are unrelated to innovations in the economy make it an ideal candidate for studying the effects of policy news.

In the main application of this paper I study the response of asset prices to election shocks. In particular, I focus on prices in the energy, clean energy, and defense sectors -- three sectors that I argue have consistently benefited (relatively) from the policies of either the Republican or Democratic party. I estimate local projections of the shock series on the log returns of sectoral stock market indexes, leveraging trade volumes in the election betting markets to weight shocks associated with noteworthy events more. Election shocks favoring Republicans immediately and persistently increase the asset prices of the energy and defense sectors, while decreasing the asset prices of the clean energy sector. The reverse is true for shocks favoring Democrats. These results provide evidence that investors are anticipating and pricing in future policy shifts driven by changes to election probabilities and election outcomes. 


To reinforce the baseline results, I construct a ``narrative election shock'' series, which only captures variation around significant events (e.g., debates and scandals) and election dates. Although this series is more sparse, its primary advantage is that it is more clean of noise associated with innovations during dates with low trade volumes. This also allows the leveraging of the entire time series (rather than only dates with nonzero trading volumes). The responses of asset prices to this measure are qualitatively and quantitatively similar to that of the original shock series.

As an alternative to the baseline and narrative election shocks, I also asses the response of stock prices to a measure that ignores election probabilities and only considers election outcomes. The response of clean energy prices remains qualitatively similar to that of the baseline measure, while the responses of energy and defense stocks are not persistent. These differences highlight the limitations of the alternative measure, which fails to capture anticipatory effects before election day and the degree of uncertainty resolved by each election outcome. 

Election shocks not only influence stock markets but also have tangible effects on the real economy. In an additional application, I estimate local projections of the election shocks on employment in industries associated with the energy, clean energy, and defense sectors. Oil, mining, clean energy generation, and certain defense manufacturing jobs change in the expected direction to a statistically and economically significant degree. There is considerable heterogeneity among the timing and magnitude of responses. In particular, mining and defense manufacturing employment respond contemporaneously, while oil and especially clean energy generation respond with a short- to medium-lag. The largest and most persistent effects are in oil and clean energy generation, which are arguably the most differentially impacted industries. There are no detectable effects to overall employment.

Taken together, these findings indicate that investors are forward-looking and that expectations of future policy, driven by election-related news, significantly shape financial market outcomes. By examining the effects of election shocks, this paper contributes to our understanding of how policy news, especially in a politically polarized environment, can move markets and influence the broader economy.

\vspace{5pt}
\noindent{\textbf{Related Literature.}} The findings of this paper are closely linked with the literature that takes the timing of fiscal policy shocks seriously. Notably, \citet{ramey_identifying_2011} shows that much of the response to defense spending increases happen in anticipation of policy implementation. \citet{romer_macroeconomic_2010} argue that expectations of tax changes play little role in the response of the macroeconomy relative to their actual implementation. \citet{leeper_quantitative_2012, leeper_fiscal_2013} map news of government spending and tax changes into the DSGE framework and argue that accounting for time-variation in the news process is vital. \citet{ricco_signals_2016} argues that clear policy communications sharpen the response to fiscal spending through the consolidation of market expectations. Most recently, and perhaps most related to this paper, \citet{hazell_deficits_2024} exploit high-frequency variation in asset prices immediately after the 2020 Georgia Senate runoff to estimate the effects of a budget deficit shock. Relative to these papers I provide two key insights. First, I show that not just election outcomes, but also changes in election probabilities can affect stock markets through updates to investors expectations about future government policy. Second, I shift the focus from aggregate outcomes to sectoral reallocation by showing that elections are an important source of fluctuations in sectors that are differentially affected by the industrial policy platforms of each political party.


My work also relates to the literature studying the financial and economic effects of political uncertainty. A key insight in this literature is that higher political uncertainty leads to higher stock market volatility \citep{pastor_political_2013, kelly_price_2016, goodell_election_2020}. Relative to this work, I shift the focus from stock market volatility to stock market returns and directly link the policy platforms of the political parties to sector-specific performance. The labor market application in this paper also closely relates to \citet{hassan_firm-level_2019} who find that firms react to increased political risk by reducing hiring and investment. In contrast with my findings, their political risk measure identifies little variation at the sectoral level.

Finally, from a methodological standpoint, this paper broadly contributes to the high-frequency time series identification literature. The pioneering work in this area has largely focused on identifying monetary policy shocks \citep{gurkaynak_sensitivity_2005, gertler_monetary_2015, nakamura_high-frequency_2018, miranda-agrippino_transmission_2021, bauer_alternative_2023, bauer_reassessment_2023, bianchi_threats_2023}. Recent work has extended these methods to study oil supply shocks \citep{kanzig_macroeconomic_2021, degasperi_identification_2023, polat_oil_2025}, uncertainty shocks \citep{ferrara_what_2018}, and Treasury supply shocks \citep{phillot_us_2025}. Closer in spirit to this paper is \citet{drechsel_estimating_2024} who studies the effects of ``political pressure'' shocks to the Federal Reserve on macroeconomic conditions. To my knowledge, no existing work has utilized election betting data to identify election shocks.

The remainder of the paper proceeds as follows. Section~\ref{sec2} discusses ``politicized sectors,'' i.e., sectors that are particularly affected by the different policy platforms of the two major U.S. political parties, and provides some preliminary evidence that the stock markets of these sectors react to election news. Section~\ref{sec3} delineates the construction of the election shock series and evaluates it by studying its movements near narrative events. Section~\ref{sec4} presents the effects of the election shocks on financial and labor markets. Section~\ref{sec5} concludes.

\section{Politicized Sectors}
\label{sec2}

\subsection{Industrial Policy and Sectoral Partisanship}
\label{sec21}

Although the particular policy platforms of the Republican and Democratic parties have evolved over the past several decades, and vary substantially at the candidate-level, the industrial policy stances of each party relative to the other have remained consistent in several aspects. I opt to focus on three sectors — energy,\footnote{Throughout, energy refers to the ``traditional'' energy sector comprised mostly of oil and coal production, while clean energy refers to the production of energy by alternative sources that do no produce emissions, i.e., nuclear, hydroelectric, solar, wind, and geothermal energy.} clean energy, and defense — which, throughout my sample period (2000–2024), have been differentially affected by the two parties. I claim that, \textit{relative to the Democratic party}, the Republican party's proposed and implemented policies to propagate the expansion of the energy and defense sectors and contraction of the clean energy sector. In this section I provide narrative and empirical evidence in support of this claim.

Starting with the energy and clean energy sectors, first note that in many cases policies that encourage production of the former disfavor the latter and vice-versa. To illustrate, consider the implementation of carbon taxes, which lowers the demand for coal and oil while increasing the demand for clean energy sources \citep{aldy_case_2013, hafstead_calculating_2017, macaluso_impact_2018}. This directly follows from the fact that carbon taxes increase the cost of consuming coal and oil, and clean energy sources are substitutes for these traditional energy sources. Similarly, \cite{acemoglu_environment_2012} argue that intervention in ``dirty'' sectors leads to more technological innovation in ``clean'' sectors. However, recent impactful bills may affect the production and technological innovation of both sources of energy in the same direction, e.g., the IRA created tax credits clean energy generation while extending and increasing credits for industrial carbon capture.\footnote{Carbon capture are technologies used to capture C02 emissions from industrial power plants. Many large oil companies like Exxon Mobil and Shell have recently invested in these technologies.} Despite these caveats, for the majority of legislation discussed in this section one of the two sectors is a clear relative beneficiary.

Table~\ref{taba1} provides a comprehensive list of significant legislation passed by recent Republican and Democratic administrations that directly affected the energy or clean energy sector. In each case a brief description of the relevant policies are included. Some notable Republican policies are the Bush administration's tax credits for refined coal production under the American Jobs Creation Act, the Trump administration's reversal of permit pipeline denials and vehicle C02 emissions standards, and Trump's withdrawal from the Paris Climate Agreement (PCA) in 2017 and 2025. Note that, while the PCA does not contain any formal constraints, it legally binds participating countries to share progress on reducing emissions, and features a commitment from developed countries to provide financial and technological assistance to developing countries. Thus, a country entering or withdrawing from the PCA serves as a signal about its subsequent energy approach.

Significant Democratic legislation has conversely leaned towards favoring the clean energy sector and disfavoring the energy sector, at least relative to Republican administrations. Several bills have inhibited production of coal and oil, such as the Obama administration's banning of new coal mining leases, and the Biden administration's raising of royalties to drill on federal lands. Each administration also passed notable legislation intended to directly increase clean energy production. For instance, Obama's American Recovery and Reinvestment Act and Biden's IRA have contributed an estimated \$90 billion (\$265 billion) in clean energy investment and tax incentives. Finally, the Obama and Biden administrations approved the construction of fewer new gas pipelines than the Bush and Trump administrations. In particular, from 2001-2024 Republican administrations approved an average of 180 percent more miles of pipe for major pipeline projects per year than Democratic administrations. Details are plotted in Figure~\ref{figa1}.

Republican (Democratic) administrations have at times enacted policies favoring clean energy (energy), such as Bush’s 2005 tax credits for solar and nuclear investment and Biden’s 2022 expansion of carbon capture incentives. However, these measures often appear as secondary provisions within broader bills that heavily invest in the opposite sector—for example, Biden’s carbon capture credits were part of the IRA, which primarily supported clean energy. While we cannot know how the opposing party would have legislated under similar conditions, some evidence suggests they may have pursued stronger policies in their preferred sector. For instance, upon taking office, Obama lifted a \$2,000 cap on Bush’s solar tax credit—suggesting an Energy Policy Act of 2005 signed by Obama might have leaned more pro-clean energy. With these caveats in mind, the analysis in Section~\ref{sec4} assumes only that firms in each sector expect to benefit \textit{more} under one party than the other.

Next I discuss the impact of each political party on the defense sector. Throughout, the defense sector refers to aerospace, naval, armored vehicle, and military weapon manufacturing as well as mission-support services, mobile communication, and surveillance and renaissance technologies. Similar to oil and coal firms' reliance on leases and permits from the federal government, by the nature of their products, companies in the defense sector heavily rely on contracts with the federal government. One parsimonious way to determine which of the political parties have acted more favorably towards defense contractors is to simply study the federal government's defense budget under each administration. 

\begin{figure}[!ht]
\caption{U.S. Defense Spending by Fiscal Year}\centering
\par
\includegraphics[width=0.6\linewidth]{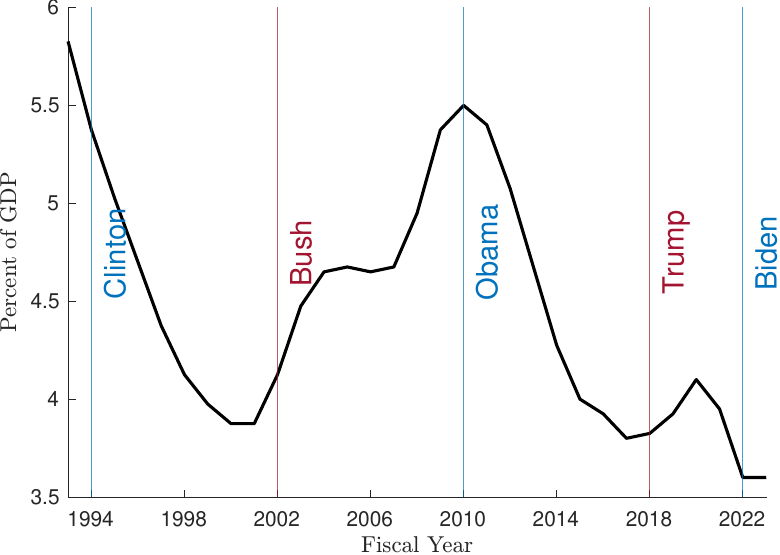} 
\begin{minipage}{\textwidth}
\vspace{5pt}
\footnotesize \textit{Notes:} This figure plots the share of U.S. GDP that is comprised of national defense spending for each fiscal year. Vertical lines indicate years when there is a change in administration. Source: U.S. Bureau of Economic Analysis.
\end{minipage}
\label{fig1}
\end{figure}

Figure~\ref{fig1} plots U.S. defense spending as a percentage of GDP from 1993 to 2024 fiscal years (FY).\footnote{Fiscal years in the federal government start in October 1st of the preceding year and end September 30th, e.g., FY2022 starts October 1, 2021 and ends September 30, 2022.} Vertical lines indicate the first fiscal year that the defense budget was set by a particular administration. Notably, from start to end of term, defense spending increased under each Republican administration and decreased under each Democratic administration. More specifically, the budget increased substantially from 3.9 to 5.4 percent from FY2001 to FY2009 under the Bush administration, by FY2017 had decreased to 3.8 percent under the Obama administration, increased modestly to 4 percent by FY2021 under the Trump administration, and finally decreased modestly again under the Biden administration to 3.6 percent by FY2024. The large increase in defense spending during the Bush administration was at least partially driven by the wars in Afghanistan and Iraq, while declines during the Obama administration are linked with the significant withdrawal of U.S. troops from these countries. It is difficult to disentangle the exogeneity of these events, e.g., would the Iraq War have taken place under alternative leadership? However, there is significant evidence provided by a collection of speeches from this period as well as the political science literature that indicates under an administration led by Al Gore, Bush's 2000 general election opponent, the war would not have taken place \citep[see for example][]{gore_iraq_2002, harvey_president_2012}.\footnote{Similarly, during the 2008 presidential elections Obama's general election opponent, John McCain criticized Obama's proposed 16 month deadline for withdrawal of troops in Iraq and offered a longer timeline \citep[for a brief overview see][]{cooper_iraq_2008}.}

\begin{figure}[htbp]
\caption{Campaign Contributions by Sector}\centering
\par
\includegraphics[width=0.8\linewidth]{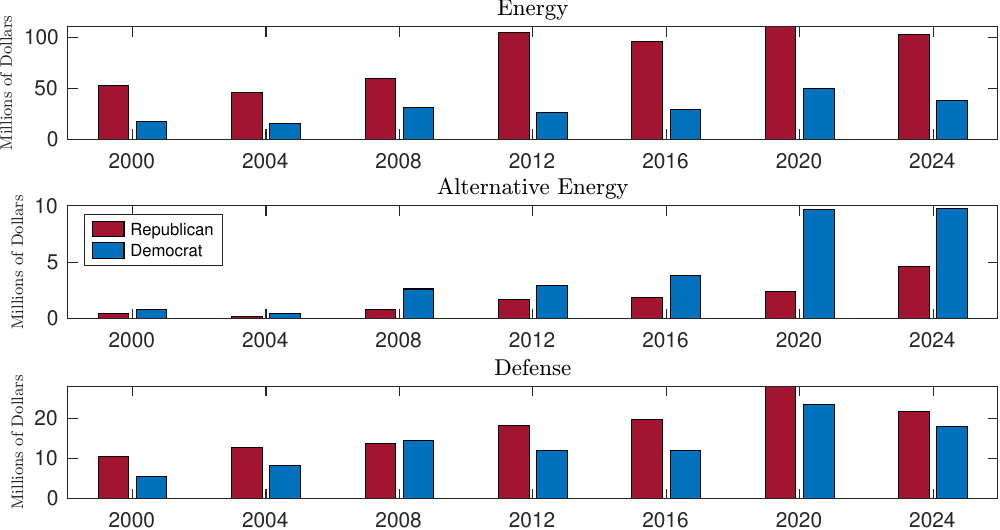} 
\begin{minipage}{1.0\textwidth}
\vspace{5pt}
\footnotesize \textit{Notes:} This figure plots the contributions of the Energy, Alternative Energy, and Defense sectors to both the Republican and Democratic party for each election cycle. Source: OpenSecrets.org.
\end{minipage}
\label{fig2}
\end{figure}

Aside from studying passed legislation and budgets, to study which sectors of the economy seek to gain more from one of the two political parties one can observe their lobbying activities. To the extent that firms' donations are driven by the proposed policies of candidates as opposed to other factors (e.g., social values), these donations reveal their monetary preferences. Figure~\ref{fig2} plots the contributions from the energy, alternative energy, and defense sectors to each political party during the 2000-2024 presidential elections. On aggregate, energy firms consistently donate the most between the three sectors, though there is significant heterogeneity between election years. There is a clear transition between 2008 and 2012 where Republican contributions nearly doubled from \$60 million to \$105 million. This discontinuity is likely driven by \cite{citizens_2010}, which unrestricted independent expenditures of corporations for political campaigns. Nonetheless, firms in the energy sector donated more than twice as much to Republican candidates compared with Democratic candidates in each election, indicating a clear preference.\footnote{The energy sector also includes clean energy firms, though they make up a small enough proportion contributors that it does not significantly bias contributions to Democratic candidates upwards.} The alternative energy sector, which includes clean energy and bioenergy firms, displays bias in the opposite direction, albeit with substantially lower aggregate contributions. Aside from 2012, firms in this category donated more than twice as much to Democratic candidates as Republican candidates. Note that the contributions of bioenergy companies lean Republican\footnote{For instance, of the top 10 alternative energy contributors in the 2024 election cycle, 2 were bioenergy companies and each donated significantly more to Republican candidates than Democratic candidates.} and thus the bias in donations towards the Democratic party in the alternative energy serves as a lower bound for that of the clean energy sector. Finally, firms in the defense sector favored Republican candidates in all but the 2008 election, with approximately 34 percent more donations to the Republican party on average. The milder differences in contributions in this sector may be partially driven by the fact that defense companies heavily rely on contracts with the federal government, i.e., it is advantageous for these companies to lobby with both parties. Indeed, in the 2024 election the top 5 defense contributors each donated at least \$500,000 to both political party. 

\subsection{Election Events and Stock Markets}
\label{sec22}

Having established the relative political preferences of these sectors, a question of interest is whether sectoral stock markets react to election events. Events that provide news about the probability that one of the two political parties will gain political power also provide news about future industrial policy. To illustrate, consider the release of negative information about ``Candidate A,'' which reduces her probability of winning the election and increases the probability of ``Candidate B.'' Suppose that Candidate B has proposed increasing taxes on the production of oil, while Candidate A denounces this proposal. Then the increase to Candidate B's probability can be viewed as a negative shock to the expectation of future oil production, the profits of oil producing firms, and even the supply of other goods produced by firms that use oil as an input in their production process. All else equal, this innovation also serves as a positive shock to the expectation of future demand for oil substitutes, e.g., nuclear energy. There is substantial evidence that stock markets react to signals about future monetary policy \citep{swanson_measuring_2021, del_negro_forward_2023}, fiscal policy \citep{baker_policy_2019}, and firm-level political risk \citep{hassan_firm-level_2019}. This suggests that, to the extent that investors observe and interpret election news as containing signals about future industrial policy, the asset prices of affected sectors will respond.

\begin{figure}[htbp]
\caption{Sectoral Stock Prices and 2024 Election Events}\centering
\par
\includegraphics[width=0.6\linewidth]{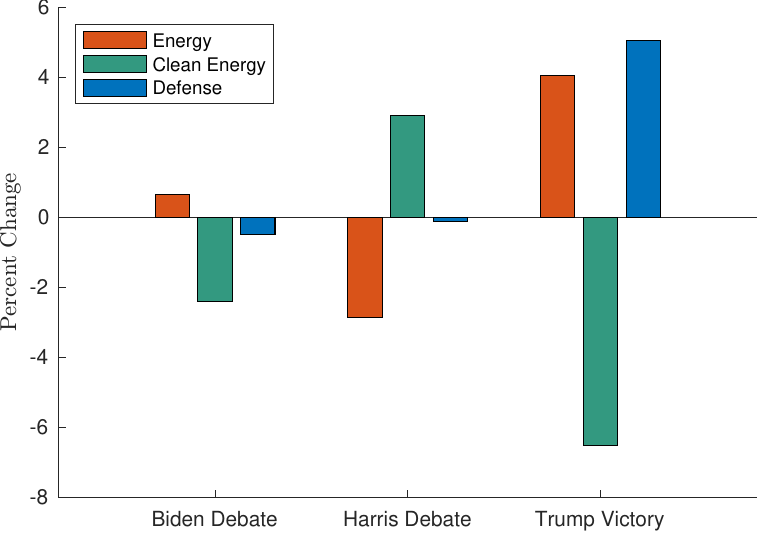} 
\begin{minipage}{1.0\textwidth}
\vspace{5pt}
\footnotesize \textit{Notes:} This figure plots the percentage change in the S\&P 500 Energy, Global Clean Energy, and Aerospace \& Defense Select indices between the day before and the day after selected 2024 election events.
\end{minipage}
\label{fig3}
\end{figure}

Figure~\ref{fig3} provides preliminary narrative evidence of this mechanism. Each bar plots the percentage change in the aggregate asset price of a given sector between the day before and the day after a significant election event. Prices are given by the relevant S\&P subindices.\footnote{The Bloomberg tickers for energy, clean energy, and defense are SPN, SPGTCED, and SPSIAD, respectively.} I consider three election events from the 2024 election: the June 27th debate between Joe Biden and Donald Trump, the September 10th debate between Kamala Harris and Donald Trump, and Donald Trump's presumptive election victory on November 5th. The first of these events provided a negative signal about Biden's chances at reelection, and was concurrent with a moderate decrease of approximately 2.4 percent in clean energy prices. The Harris debate, for which Harris was the consensus winner, improved the Democrats' chances at the White House and energy and clean energy prices changed by -2.9 and 2.9 percent, respectively. Finally, Trump's victory on election day is associated with substantial contemporaneous increases to energy and defense prices of 4 and 5 percent, and a decrease in clean energy prices of 6.5 percent.

These price movements are suggestive, but a more formal analysis is required to understand the extent to which stock markets respond contemporaneously to innovations in election probabilities. It is possible that these movements are particular to the 2024 election, perhaps due to the intricacies of Biden, Harris, or Trump's policy platforms. Furthermore, it is important to consider whether any stock market effects are persistent. \citet{kwon_extreme_2025} argue that investor beliefs overreact (underreact) to extreme (subtle) types of corporate news leading to reversal (drift) in asset prices. Thus the short-term reactions suggested by Figure~\ref{fig3} may only capture the overexcitement of investors. Finally, does election news have any real effects? If firms update their expectations the same way that investors do, they may alter their hiring or production decisions before any bills are passed, or even before the inauguration of the next president. The stock market performance of a firm may also indirectly affect its output, e.g., after an increase to asset prices, financially constrained firms are more likely to raise capital through the equity-financing channel \citep{baker_when_2003, campello_stock_2013}. In the next section I construct an election shock series to study these possibilities.

\section{Election Shocks}
\label{sec3}

\subsection{Data}
\label{sec31}

The primary data used to construct the election shocks come from Iowa Electronic Markets (IEM), a futures contract market operated by the University of Iowa for research purposes. IEM publicly provides data on its winner take all futures betting markets for presidential races. Betting markets have been shown to more accurately predict elections than polling data \citep{berg_prediction_2008}. Relative to other betting markets, IEM has several desirable properties: (i) bets are capped at \$500, mitigating the influence of single traders; (ii) the markets have consistently high trading volumes, especially in the last few months of an election; (iii) the data are at a daily frequency, which is vital for studying stock markets; and (iv) market data is available for every presidential election since 1988\footnote{I focus on elections starting in 2000 due to the lack of availability of trading volumes and sectoral price indices for earlier elections.}, and in all election years data are available starting in June or earlier and extend through election day in November. One drawback of these data are that the payouts are based on which party wins the \textit{popular vote} rather than the electoral college. In other words, the implied probability of a candidate winning based on these futures prices are biased if her chances of winning the popular vote are different from her chances of winning the election. Indeed, this is particularly relevant for recent elections, e.g., based on other betting markets Harris's predicted probability of winning the popular vote was much higher than her probability of winning.\footnote{For example, odds from Polymarket.com the day before the election implied Harris had a 71 percent chance to win the popular vote, but a 41 percent chance to win the presidency.} While the last three elections had electoral college bias in favor of the Republican candidates, there is no clear historical systematic bias \citep{erikson_electoral_2020, coleman_electoral_2023}. With this in mind, to construct the election shocks I rely on the implicit correlation between innovations to both a candidates' probability of winning the popular vote and the probability that they win electoral college. 

\begin{figure}[htbp]
\caption{2024 IEM Implied Election Probabilities}\centering
\par
\includegraphics[width=0.8\linewidth]{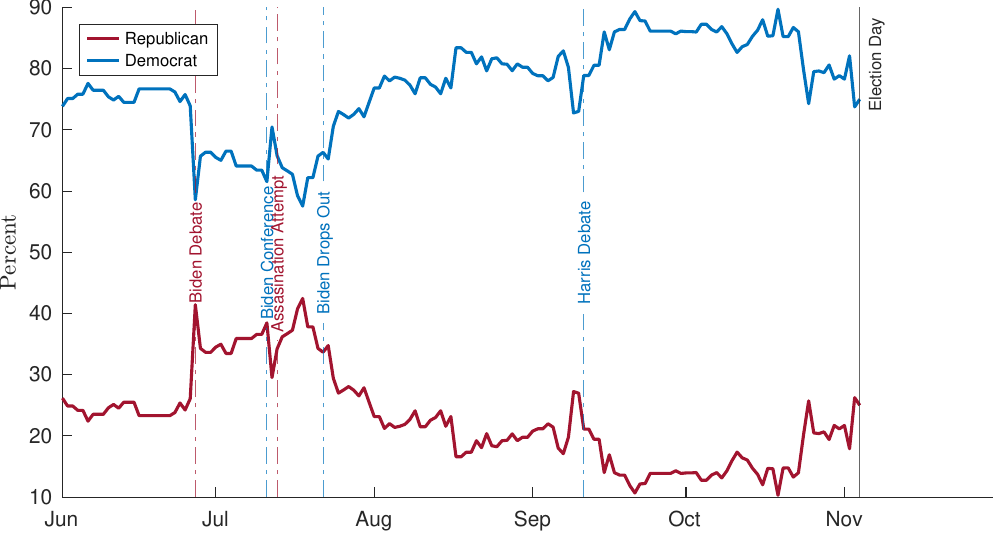} 
\begin{minipage}{1.0\textwidth}
\vspace{5pt}
\footnotesize \textit{Notes:} This figure plots the election probabilities implied by the prices of the IEM winner take all futures contracts for the 2024 presidential election. Vertical lines indicate the dates of significant election events.
\end{minipage}
\label{fig4}
\end{figure}

Figure~\ref{fig4} plots the 2024 Republican and Democratic candidates' implied winning probabilities from June 1st through November 4th, the day before the election. The probabilities sum to 100, i.e., an increase to the Republicans' probabilities are associated with a 1-to-1 decrease in the Democrats' probabilities. Democratic candidates were each favorites to win the popular vote throughout the sample, although there is significant heterogeneity in the magnitude of this advantage. Vertical lines indicate notable events that are associated with substantial movements in the data. The first and most notable event is the aforementioned debate between Biden and Trump, which is associated with a contemporaneous 15pp increase to Trump's probability of winning, albeit leveling off roughly 10pp higher than pre-debate levels. On July 11, Biden held a news conference that received mixed reactions, but overall was seen as a positive signal relative to his debate performance. This event is associated with a moderate increase to the Democrats' odds of presidency. Just two days later, Trump survived an assassination attempt during a rally in Pennsylvania, which bolstered his election probabilities. A medium-term increase in the Democrats likelihoods appears to be linked to Biden dropping out of the race and endorsing Harris on July 21. Finally, the aforementioned debate between Harris and Trump is associated with a steady increase in the Democrats' odds, which peak about 16pp higher roughly two weeks after the debate. See Figure~\ref{figb1} for the probabilities of other elections in the sample and Table~\ref{tabc2} for information on other identified events. Overall, these data appear to adequately track fluctuations in each party's election probabilities around significant events.

Aside from the election betting data, I obtain daily asset prices from Bloomberg Terminal, financial and macroeconomic data from FRED, and vintages of macroeconomic data from ALFRED. Finally, in Section \ref{sec43} I use industry-specific employment data obtained from the Quarterly Census of Employment and Wages (QCEW), which are available from January 2001 through March 2024.

\subsection{Construction of the Election Shocks}
\label{sec32}

The goal of this section is to construct a shock series that captures the types of fluctuations in election probabilities that are emphasized in Figure~\ref{fig4}. However, a significant challenge to this exercise is the possibility that movements in election probabilities also reflect changes in voter preferences given changes to the state of financial markets or the real economy. For example, consider an election year in which the economy is experiencing high inflation. Voters may blame the incumbent party, lowering their chances at reelection. Alternatively, voters may trust one of the political parties more (regardless of incumbency) to handle inflation. With the goal of estimating the effects of changes in election probabilities on stock markets, these possibilities introduce severe endogeneity concerns. Two potential approaches to mitigate this identification threat are (i) taking a narrative approach by selecting changes around events plausibly exogenous to the economy; and (ii) orthogonalizing the election probability movements to developments in the economy. Relative to the first approach, the second approach has the advantage of producing a less sparse series, and the drawback of introducing more noise. In this section I take the second approach, returning to the narrative approach in Section \ref{sec42}.

Consider the following linear\footnote{In unreported results I instead use a fractional regression à la \citet{papke_econometric_1996} and obtain similar results} equation:
\begin{align}
    \pi^{\text{R}}_{t} & = \alpha_0 + \sum_{s=1}^{5}\alpha_{1, s}\pi^{\text{R}}_{t-s} + \sum_{s=0}^{5}\alpha_{2, s}\textbf{X}_{t-s} + \alpha_3 \text{Pres}^{\text{R}}_{t} + \sum_{s=0}^{5}\alpha_{4, s}'\left( \textbf{X}_{t-s} \times \text{Pres}^{\text{R}}_{t}\right) + \varepsilon_t,
    \label{eq1}
\end{align}
where $\pi^{\text{R}}_{t}$ denotes the probability that the Republican candidate wins the next election as of day $t$, $\textbf{X}_{t}$ is a vector of contemporaneous economic series, and $\text{Pres}^{\text{R}}_{t}$ is dummy variable equal to 1 (0) when the current president is a Republican (Democrat). A week (i.e., 5 business days) of autoregressive lags are included to purge the series of serial correlation. The remaining series represent potential sources of variation that may affect economic outcomes as well as election probabilities. For example, the set of interaction terms may affect election probabilities through the voters (dis)pleasure with the incumbent party, and affects financial and real conditions if each party reacts differently to economic conditions. If a sufficient set of predictors are included in $\textbf{X}_{t}$, then $\varepsilon_t$ identifies the variation in $\pi^{\text{R}}_{t}$ that comes from sources other than developments in the economy. Thus, it is important to carefully choose the content of this vector so that one may interpret $\varepsilon_t$ as an exogenous shock. In my baseline specification I set:

\begin{align}
    \textbf{X}_{t} \equiv \left[\Delta i_t,\, \Delta \text{sp}500_t,\, \mathbbm{1}_e(t) \cdot \Delta \text{emp}^{1m}_t,\, \mathbbm{1}_c(t) \cdot \Delta \text{cpi}^{1m}_t,\, \mathbbm{1}_i(t) \cdot \Delta \text{ind}^{1m}_t  \right]', 
\end{align}
where $\Delta i_t$ and $\Delta \text{sp}500_t$ are the 1-day percentage change in 2-year Treasury yields and the S\&P 500 and $\Delta \text{emp}^{1m}_t$, $\Delta \text{cpi}^{1m}_t$, and $\Delta \text{ind}^{1m}_t$ give the most recent 1-month percentage change in employment, the CPI, and industrial production. $\mathbbm{1}_e(t)$, $\mathbbm{1}_c(t)$, and $\mathbbm{1}_i(t)$  are indicator functions equal to 1 \textit{on the day of the data release} for each respective series.\footnote{The employment situation from the preceding month is typically released on the first Friday of each month, while CPI and industrial production are released with roughly a 2-week lag.} Thus, the last three components of $\textbf{X}_{t}$ take on values of 0 each day of the month except days where new information about the relevant macroeconomic series is released. It is critical to take the timing of these series' release seriously since traditional stock markets (and plausibly election betting markets) react to news about the economy in real-time \citep{boyd_stock_2005, lapp_impact_2012}. It is also important to account for data revisions to the initial releases to accurately reflect the information available at a given time. Therefore, for each component of $\textbf{X}_{t}$, I use values from the data vintages corresponding to each specific date, rather than relying on the most recent vintage, which introduces look-ahead bias. Note that equation~(\ref{eq1}) includes a week of lags for the base and interaction terms of $\textbf{X}_{t}$, which allows for a delayed market reaction.\footnote{In unreported results I exclude the indicator functions so as to treat stale and new information the same. Results in Sections~\ref{sec3} and \ref{sec4} are unaffected.} These components can be interpreted similarly to the ``economic news'' measure used by \citet{bauer_alternative_2023, bauer_reassessment_2023} to orthogonalize monetary policy shocks to information released between the Blue Chip Economic Indicators survey and FOMC announcements. One key difference between these two applications is the frequency of data -- the monetary policy shocks are at the FOMC announcement frequency as opposed to the daily frequency. 

I estimate equation~(\ref{eq1}) on the data from the 2000-2024 elections using OLS. Note that in non-election periods $\pi^{\text{R}}_{t}$ is unavailable, so these dates are excluded from the estimation.\footnote{In practice this results in roughly 3-year time gaps between the end of an election and the start of the subsequent one. 5 observations are lost at the start of each election period as a result of the lag structure of the model. An alternative way to view this data is as an unbalanced panel with 7 groups.} Results are reported in Table~\ref{tabc1}. To summarize, the R-squared is 0.885, indicating that the model fits the data well. The first two auto-regressive lags are unsurprisingly important as the election probabilities are highly serially correlated. The other terms contain little predictive content. See Appendix~\ref{secc} for more details.

Finally, I denote
\begin{align}
    \text{Shock}_t = \hat{\varepsilon}_t,
\end{align}
where $\hat{\varepsilon}_t$ represents the remaining 11.5 percent of variation in the data not explained by the model, which I interpret as an exogenous election surprise throughout the remainder of the paper. 

\subsection{Analysis of the Election Shocks}
\label{sec33}

\begin{figure}[h!]
\caption{Election Shocks}\centering
\par
\includegraphics[width=0.9\linewidth]{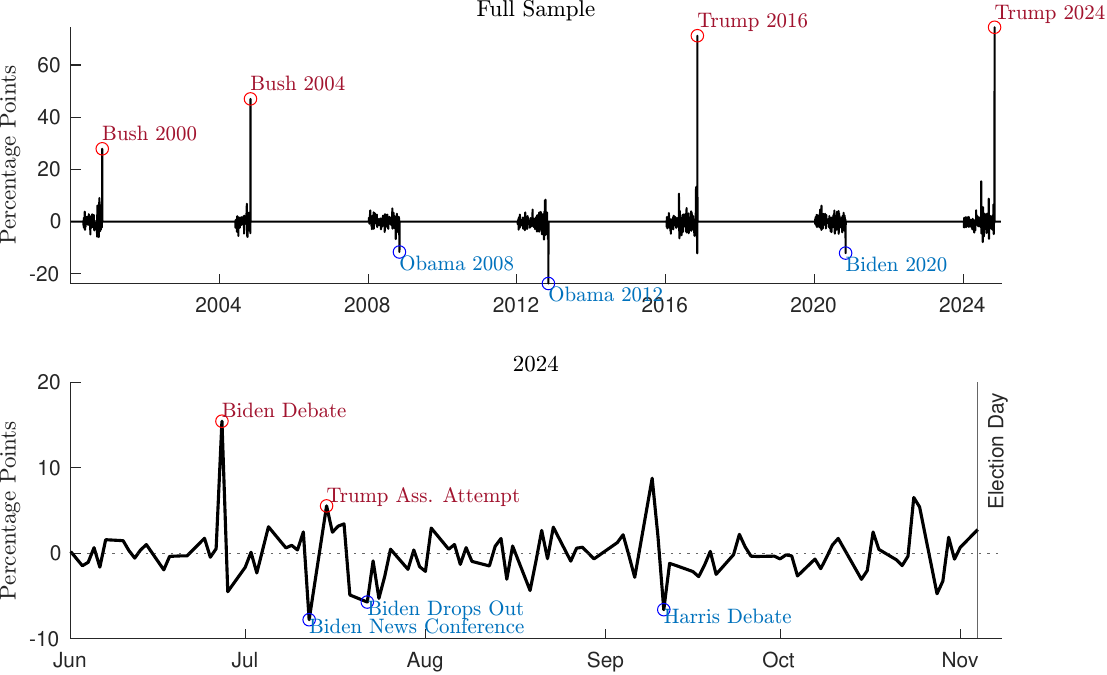} 
\begin{minipage}{1.0\textwidth}
\vspace{5pt}
\footnotesize \textit{Notes:} The top panel of this figure plots $\text{Shock}_t$ over the full sample. The bottom panel zooms into the 2024 election period for the same series. Positive values represent exogenous changes in election probabilities in favor of the Republican candidate. Values outside of election cycles are set to zero. In the top panel red and blue circles indicate election days. In the bottom panel red and blue circles indicate significant election events that are associated with shocks in favor of the Republican and Democratic candidate, respectively.
\end{minipage}
\label{fig5}
\end{figure}

The first panel of Figure~\ref{fig5} plots $\text{Shock}_t$ over the entire sample. Positive shocks are associated with surprise increases to the Republicans' probability of winning the upcoming election, while the converse is true for negative shocks. Shocks are set to equal 0 outside of election periods. The largest shocks are generally associated with election day, when $\pi^{\text{R}}_{t}$ goes to 0 or 1.\footnote{There was still some uncertainty in the days after the 2000 and 2020 elections, but in my baseline specification I ignore this complication due to data limitations. The 2000 election poses a larger difficulty due to a month-long lawsuit over the outcome. In unreported results I exclude the 2000 election and the main results of this paper are largely unchanged.} However, there is substantial heterogeneity in the size of these shocks. For instance the shocks associated with the Obama 2008 and Biden 2020 victories are each approximately -12pp, while the shock associated with the Trump 2016 victory is a staggering 71pp. In an informal sense, the 2008 and 2020 results were closer to a resolution of uncertainty than a ``shock,'' as Obama and Biden were widely expected to win. On the other hand, the 2016 result was a true surprise given the predictions of political analysts and media outlets. Also noteworthy, the magnitude of the shock associated with Trump's 2024 victory is slightly overstated given the tightness of the race leading up to election day. This is a result of the original series measuring the odds of each candidate winning the popular vote, rather than the electoral college.\footnote{In Section~\ref{sec4} this overstatement biases the magnitude of my results downward as the sizable stock market activity following the election are associated with a smaller surprise than the one suggested here.} Overall, the shocks associated with election days realistically differentiate between the degree of surprise associated with each result. This feature of the series is critical, as stock markets are likely to react differently to expected and unexpected election outcomes.

The second panel of Figure~\ref{fig5} focuses in on the election shock series for the 2024 election period. Positive shocks are associated with the Biden debate and Trump assassination attempt, while negative shocks are associated with the post-debate news conference, Biden dropping out of the race, and the Harris debate. In several cases, these events are followed by a cluster of movements in the election odds as opposed to one large contemporaneous jump. One interpretation of these patterns is that the event itself provides a noisy signal to traders, and in the days following more information continues to disseminate. For example, there are three consecutive positive shocks following the Trump assassination attempt, which may be attributed to traders' responses to developments in the investigation, media coverage, and Trump's first public appearance at the Republican National Convention following the attempt. A similar interpretation is that betting markets respond initially, but prices continue to adjust as information about voter responses come in (i.e., the release of polls). Nonetheless, the series is able to capture plausibly exogenous movements in the candidates' election probabilities throughout the 2024 election cycle. Figure~\ref{figc1} plots the shocks from each election cycles in my sample separately. Interestingly, there is considerable heterogeneity in the variance of the series between election years. Aside from the Biden debate and the realization of several election outcomes, the largest values are associated with the October 28, 2016 announcement by former FBI director James Comey regarding the reopening of an investigation on leaked emails by Hillary Clinton from her time as the U.S. Secretary of State. In all cases the series moves near notable election events in the expected direction. Section \ref{sec42} discusses these events further.

\section{The Sectoral Effects of Election Shocks}
\label{sec4}

\subsection{Stock Markets}
\label{sec41}

Having constructed the election shock series, I return to study the questions motivated in Section \ref{sec22}. First, I consider daily local projections of this series on the log-returns of sectoral asset prices. The baseline empirical specification is represented by the following equation:
\begin{align}
    \Delta y^i_{t+h} = \beta^i_h + \gamma^i_h \text{Shock}_t + \delta^i_h \Delta y^{i, 1m}_{t-1} + \eta^i_{t+h}.
    \label{eq4}
\end{align}
where $y^i_t$ is the log price of a sector stock index, $\Delta y^i_{t+h}$ is an $h$-day ahead long difference, $\Delta y^{i, 1m}_t$ is the one-month change in $y^i_t$, and $\gamma^i_h$ are the coefficients of interest. Regressions are specified in differences to reduce bias in the estimates of $\gamma^i_h$ per the recommendations of \citet{piger_differences_2023} \citep[also see][]{jorda_local_2025}. For each sector and horizon I estimate equation~(\ref{eq4}) using weighted OLS, where the weight is given by the IEM trade volumes on day $t$.\footnote{Trade volumes are plotted in Figure~\ref{figb2}. Energy and defense prices are available for the entire
sample, while clean energy prices are available starting in November 2003. Additionally, I exclude clean energy prices from September 2008 through December 2008 in the estimation due to a near 60 percent decrease during this period associated with the financial crisis. In unreported results I include these dates and point estimates are largely unchanged.} In general, weighting by trade volumes reduces the variance of the estimates of $\gamma^i_h$, since election probability movements on days with low trade volumes may be driven by noise. In other words, it is advantageous to place more emphasis on observations with high trade volumes since these observations are associated with significant election events and dates closer to election day when markets are more tuned into election news.\footnote{One drawback of this approach is that take place in non-election periods are excluded from the estimation. In Section \ref{sec42} I explore an alternative approach that circumvents this weakness.} An identification concern with examining far out responses is that changes in election probabilities are correlated with future political power, which in turn may result in the signing of legislation with direct effects on stock markets. To avoid this challenge I focus on the responses only up to 65 business days (3 months) after the shocks, which is approximately the distance between election day and inauguration.

\begin{figure}[!ht]
\caption{Stock Price Response to a 10pp Election Shock}\centering
\par
\includegraphics[width=\linewidth]{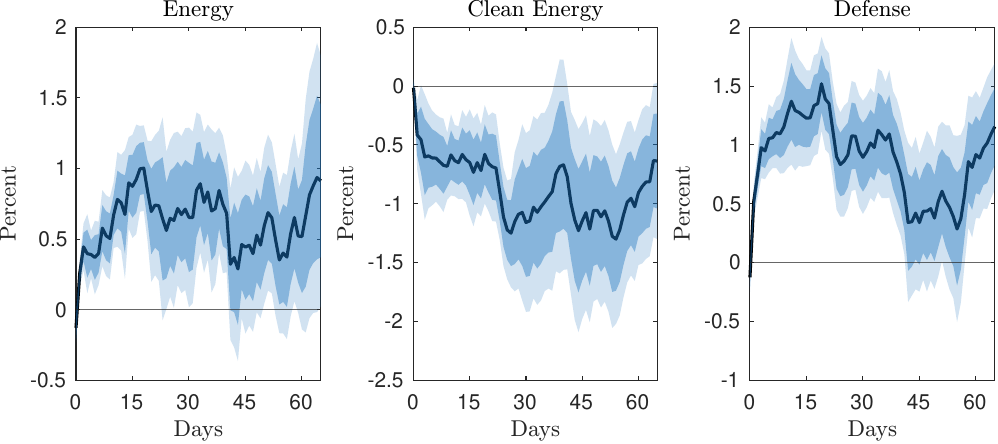} 
\begin{minipage}{1.0\textwidth}
\vspace{5pt}
\footnotesize \textit{Notes:} This figure plots the estimated impulse responses of sectoral stock prices to an election shock, normalized to increase the Republican's probability of winning by 10pp. Solid black lines report the point estimates, while the dark and light shaded areas give the 68 and 90 percent confidence intervals, respectively. 
\end{minipage}
\label{fig6}
\end{figure}

Figure~\ref{fig6} plots the daily response of the energy, clean energy, and defense asset prices to a 10pp ``Republican Shock,'' i.e., an election shock that produces a contemporaneous 10pp increase to the the Republican candidate's probability of winning the election.\footnote{Figure~\ref{figd1} plots the response of $\pi^{\text{R}}_{t}$. Overall, election shocks have persistent effects on election probabilities.} The black lines are point estimates and the shaded areas represent the 68 and 90 percent confidence intervals. Standard errors are estimated using the Newey-West HAC-consistent estimator with a bandwidth of $\frac{3}{4}\text{T}^\frac{1}{3}$, where T is the number of observations.

Energy prices increase rapidly in response to the shock, peaking at about 1 percent higher than baseline levels roughly four weeks after the shock. The effect of the shock is also persistent -- at the furthest horizon prices remain significantly above baseline levels. Clean energy prices respond with a similar magnitude in the opposite direction, albeit with larger confidence intervals. Nonetheless, the response is significant in the short-term, and the point estimates remain well below baseline levels at the furthest horizon. The mirrored movements of energy and clean energy prices are reassuring -- if the election shocks were confounded with, for instance, demand shocks, the two stocks would likely move in the same direction. Finally, defense prices display a similar pattern to energy prices peaking at about 1.5 percent higher than baseline levels. Overall, these results substantiate the narrative evidence provided in Section~\ref{sec2} while providing additional insight about the medium-term effects of election news.

A crude approach relative to the one taken here is to focus only on election outcomes by replacing Shock$_t$ with a variable equal to 1 (-1) on the day that a Republican (Democratic) candidate wins an election, and 0 otherwise. Figure~\ref{figd2} plots the responses to this variable. The response of clean energy prices remains qualitatively similar, while the measure fails to capture persistent changes to defense and energy prices. This suggests that considering the entire election sample and the degree of ``surprise'' of each election outcome implied by the election probabilities is important. This is unsurprising given the anecdotal evidence provided in Figure~\ref{fig3} -- asset prices on election day already reflect market expectations.

The main results are robust to other sensible specifications. In the main specification I treat the election shock series as observed, ignoring estimation uncertainty involved in its construction when computing standard errors. Figure~\ref{figd3} plots the estimation results of a ``one-step'' alternative that estimates the response of asset prices to $\pi^{\text{R}}_{t}$ while controlling directly for the right-hand-side components of equation~(\ref{eq1}).\footnote{By the Frisch-Waugh-Lovell theorem, the point estimates  of this approach are nearly analytically equivalent to the baseline specification. They are different due to the inclusion of $\Delta y^{i, 1m}_{t-1}$ in equation~(\ref{eq4}), the use of weights, and in the case of clean energy differences in the sample between the 1st and 2nd stages.} The point estimates and confidence intervals do not meaningfully differ from the baseline, indicating the generated regressor problem is not consequential.

Two potential factors that feasibly confound the main results are irregularities in stock markets associated with the 2007-2008 financial crisis and the COVID-19 pandemic. The former of these events spanned the entire 2008 election cycle, while the latter spanned the majority of the 2020 election cycle. If rapid changes in asset prices related to one of these two events accompanied large swings in election probabilities, my estimates could be biased in either direction. To alleviate this concern I re-estimate equation~(\ref{eq4}) after dropping observations from the 2008 and 2020 elections and plot the resulting impulse responses in Figure~\ref{figd4}. Reassuringly, the results are largely unchanged.

\subsection{A High-Frequency Narrative Approach}
\label{sec42}

While the approach employed in the previous section has several advantages, including its simplicity, there are a some drawbacks as well. Election betting markets are noisy and contain a lot of price movements that may not be well explained by actual shifts in election probabilities. Additionally, even if these markets perfectly capture reality, investors and other economic agents may not react to subtle changes in the election outlook that happen on a day-to-day basis. Contrarily, investors are more likely to react to newsworthy events that lead to large changes in the election outlook. With this in mind, here I construct an alternative measure that considers only changes in election probabilities around notable narrative events, and repeat the analysis from the previous section.

Table~\ref{tabc2} provides a list of dates and associated events from the 2000-2024 elections. Each presidential and vice presidential debate is included along with other exogenous events that occur in conjunction with notable shifts in election probabilities.\footnote{Each Democratic and Republican national conventions are also included, since these events are widely televised and scrutinized by political analysts.} In practice, constructing a narrative-based shock series requires making a choice about how tight of a window around each date to attribute changes in election probabilities. It is standard in the high-frequency identification literature to use one-day or even intraday windows. For this application, it is not clear that the tightest possible window (of one day) is advantageous. As discussed in Section \ref{sec33}, in some cases, election probabilities react somewhat sluggishly to events, which is captured by the fact that many important dates are followed by a cluster of shocks of the same sign rather than one discrete movement. However, larger windows increase the likelihood of contaminating the measure with noise. With this in mind, for my baseline narrative shock, I consider a 5-day window\footnote{In unreported results I also consider 3-day and 1-day windows with no qualitative affect on the results.} and let:
\begin{align}
 \text{Shock}_t^\text{N} = \mathbbm{1}_N(t) \cdot \sum_{s=0}^{4}\text{Shock}_{t+s}, 
 \label{eq5}
\end{align}
where $\mathbbm{1}_N(t)$ is an indicator function equal to one for each the dates given in Table~\ref{tabc2}. Note that the use of the original shock measure rather than the raw election probabilities reduces the potential for confounding the narrative shocks with financial and economic ones. Figure~\ref{figc2} plots $\text{Shock}_t^\text{N}$ both for the full sample and separately for each election window. Similarly to $\text{Shock}_t$, the largest shocks tend to be on election day (particularly in 2016 and 2024), though for the majority of years the absolute sum of non-election day values is larger than the election day value. As in the baseline series, shocks in favor of Republicans and Democrats tend to align with the types of events one would expect given the narrative information around each selected date. See Appendix~\ref{secc} for more discussion.

\begin{figure}[!ht]
\caption{Stock Price Response to a 10pp Narrative Election Shock}\centering
\par
\includegraphics[width=\linewidth]{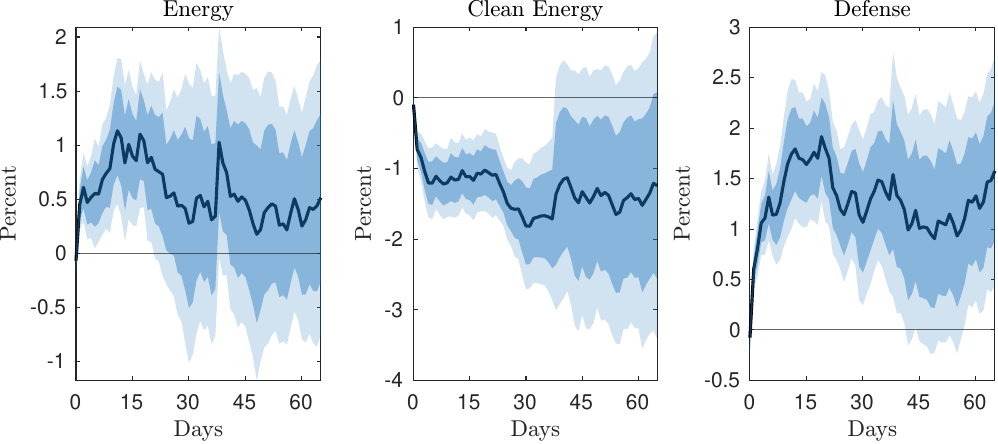} 
\begin{minipage}{1.0\textwidth}
\vspace{5pt}
\footnotesize \textit{Notes:} This figure plots the estimated impulse responses of sectoral stock prices to a narrative election shock, normalized to increase the Republican's probability of winning by 10pp. Solid black lines report the point estimates, while the dark and light shaded areas give the 68 and 90 percent confidence intervals, respectively. 
\end{minipage}
\label{fig7}
\end{figure}

With the alternative series in hand, I re-estimate equation~(\ref{eq4}), replacing $\text{Shock}_t$ with $\text{Shock}_t^\text{N}$. Figure~\ref{fig7} plots the impulse responses of each sectoral asset price to a 10pp narrative ``Republican shock.'' The results largely mimic that of Figure~\ref{fig6}. The energy response tends to be a bit smaller in the medium-term, while the clean energy and defense responses are larger, albeit with wider confidence intervals at longer horizons. Despite these differences, the estimated responses are of the same order of magnitude as in the baseline results, reinforcing the takeaways from the previous section.

\subsection{Labor Markets}
\label{sec43}

I now turn to studying whether election shocks have any real economic effects. I focus specifically on the response in labor markets, in line with \citet{hassan_firm-level_2019} who find that firms exposed to political risk alter their hiring decisions. In the context of this paper, I posit that firms are likely to react to news by investing in workers even before policy-related changes in demand are realized.
The baseline specification is:
\begin{align}
    \Delta y^i_{t+h} = \phi^i_{h} + \theta^i_h \overline{\text{Shock}}_t + \sum_{s=1}^{12}\Psi^i_{h,s}{}' \mathbf{W}^i_{t-s} + \zeta^i_{t+h},
    \label{eq6}
\end{align}
where $y^i_t$ is the log private employment of industry $i$, $\Delta y^i_{t+h}$ is an $h$-\textit{month} ahead long difference, $\overline{\text{Shock}}_t$ is the monthly sum of $\text{Shock}_t$, 
and $\mathbf{W}^i_t$ is a vector containing $y^i_{t}$, the unemployment rate, and 1-month differences in log CPI, log PCE, and log industrial production. $\theta^i_h$ are the coefficients of interest. Aside from aggregate employment, the set of industries I consider are oil drilling and extraction, mining and quarrying, clean energy generation, aerospace manufacturing, and ship and tank manufacturing. Table \ref{tabd1} provides details about the set of industries. For each horizon and industry, I estimate equation~(\ref{eq6}) on the data from January 2002 through March 2024.\footnote{I omit observations where $t+h$ is between March and December 2020 to avoid complications related to the COVID-19 period. See \citet{lenza_how_2022} for a detailed discussion of this issue.}

\begin{figure}[!ht]
\caption{Employment Response to a 10pp Election Shock}\centering
\par
\includegraphics[width=\linewidth]{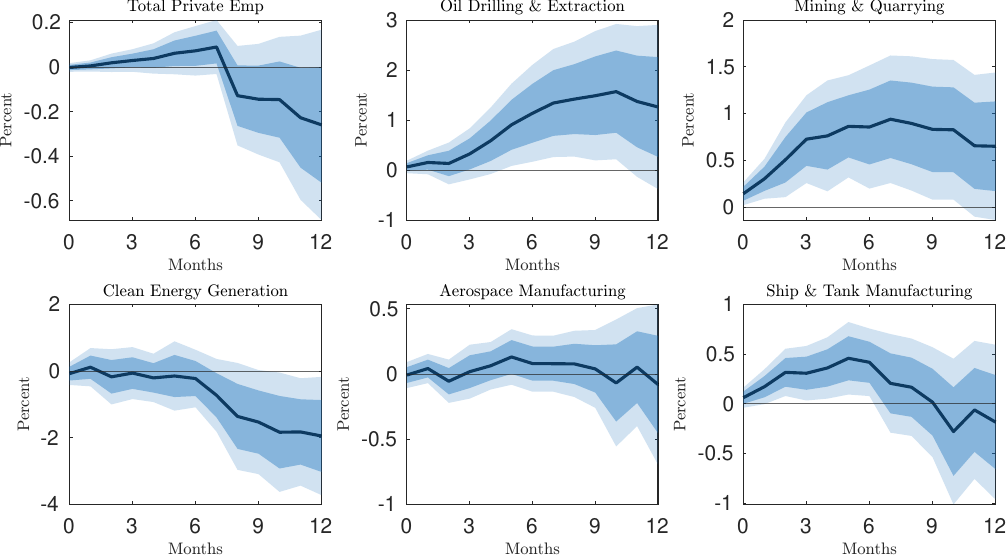} 
\begin{minipage}{1.0\textwidth}
\vspace{5pt}
\footnotesize \textit{Notes:} This figure plots the estimated employment response to of sectoral an election shock, normalized to increase the Republican's probability of winning by 10pp. Solid black lines report the point estimates, while the dark and light shaded areas give the 68 and 90 percent confidence intervals, respectively.  
\end{minipage}
\label{fig8}
\end{figure}

Figure~\ref{fig8} plots the estimated 0-12 month-ahead industrial employment response to a 10pp election shock along with the 68 and 90 percent confidence bands.\footnote{Standard errors are estimated in the same fashion as in the previous section.} The first panel gives the total employment response, which remains close to zero and statistically insignificant throughout. 

Industries associated with the energy sector respond positively to a Republican shock. Oil drilling and extraction employment increases with a slight lag and point estimates peaking at roughly 1.6 percent 10 months after the shock. Similarly, mining and quarrying employment increases contemporaneously, with the response peaking 7 months after the shock at roughly 1 percent. On the other hand, jobs in clean energy generation display little response in the short-term but decrease by approximately 2 percent 12 months after the shock. Overall, the energy and clean energy sectors respond in the expected direction given the discussion in Section~\ref{sec2}, albeit with varying lags and magnitudes. As in the stock markets application I interpret 0-3 month effects as being driven purely by the election news channel, while responses after the 3-month mark may be driven by a combination of news and actual policy implementation.

Aerospace manufacturing, which makes up approximately 76 percent of defense-related manufacturing jobs in the sample does not respond. On the other hand, ship and tank manufacturing shows a small contemporaneous increase, albeit without lasting effects.\footnote{Figure~\ref{figd5} reveals that the effects on tank manufacturing, which makes up a relatively smaller share of jobs, are more persistent.} Taken together, there is some evidence that election shocks have an impact on defense manufacturing employment. Given the significant lobbying of defense contractors to both political parties, it is unsurprising that these effects are muted relative to jobs in energy and clean energy.

\section{Conclusion}
\label{sec5}

Studying the effects of government policies on the macroeconomy generally requires one to measure the response of relevant economic variables to the policies' implementation. However, previous work has argued that much of the response often takes place in anticipation of policy implementation, complicating researchers' task. This paper provides evidence of these anticipation effects in the context of U.S. presidential elections. Election outcomes and changes in election probabilities throughout the election period have strong effects on stock markets.  Relative to the Democratic party, changes in favor of the Republican party are correlated with future policies that expand the energy and defense sectors and shrink the clean energy sector. In response to these changes, energy and defense asset prices increase while clean energy prices decrease. Similarly, employment in industries associated with each of these sectors change in the short- to medium-term. In particular, oil, mining, and some types of defense manufacturing jobs increase in response to Republican shocks while clean energy generation jobs decrease. One potential avenue for future research is to study whether the labor market response to these shocks is driven by the stock market response, firm expectations, or another mechanism. Future work may also contribute by disentangling news-driven effects from policy implementation in the medium- and long-term.

\pagebreak\relax

 {\setlength \parindent {1.5em}%
 \setlength \parskip {0pt}%
 \baselineskip=12pt\parskip=12pt \parindent=24pt
 \bibliography{ref_edit}}

\newpage
\appendix
\doublespacing
\counterwithin*{equation}{section}
\renewcommand\theequation{\thesection\arabic{equation}}
\centerline{{\Huge Online Appendix}}

\section{Political Polarization in Energy Policy}
\label{seca}

\setcounter{figure}{0}
\renewcommand{\thefigure}{A\arabic{figure}}
\setcounter{table}{0}
\renewcommand{\thetable}{A\arabic{table}}

\begin{figure}[!ht]
\caption{Pipeline Construction Approval}\centering
\par
\includegraphics[width=0.6\linewidth]{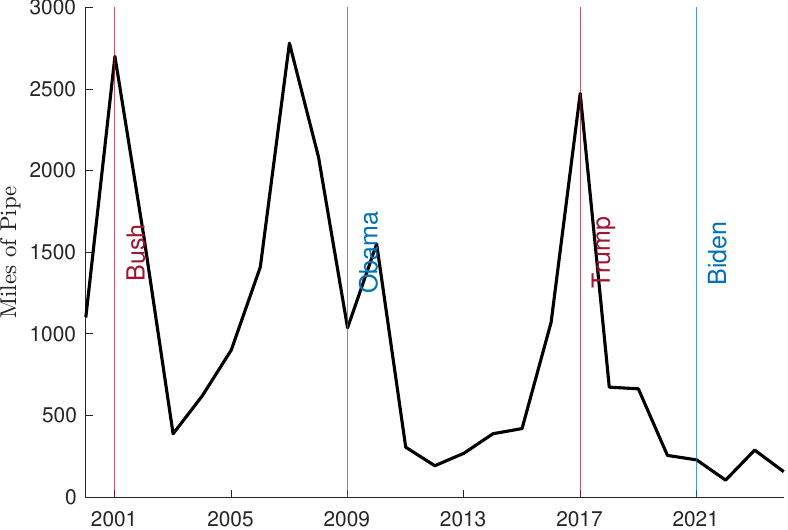} 
\begin{minipage}{1.0\textwidth}
\vspace{5pt}
\singlespace
\footnotesize \textit{Notes:} This figure plots the miles of natural gas pipeline approved by the federal government in each year. Vertical bars indicate years with administration changes. Source: Federal Energy Regulatory Commission.
\end{minipage}
\label{figa1}
\end{figure}

\doublespacing
This section provides additional information on energy policy implementation of Republican and Democratic administrations. As discussed in Section \ref{sec2}, over the past several decades Republican administrations have generally passed legislation that stimulates the expansion of oil, gas, and coal and the contraction of clean energy sources relative to Democratic administrations. One of the most direct ways that the Federal government benefits (or inhibits) the traditional energy sector is by its approval of pipeline construction. Pipelines are used to transport natural gas from production areas to storage facilities and consumers. Figure~\ref{figa1} plots the miles of pipe approved by the Federal government in each year from 2000 through 2024. Red vertical bars indicate the start of a new Republican administration, while Blue bars indicate the start of a new Democratic administration. Overall, Republicans approved 180 percent more miles of pipe on average per year throughout the sample. Notably, the start of each Republican administration is associated with a large spike in approval. In practice, this may be due to the approval of projects that were inhibited by their Democratic predecessors. For instance, within Trump's first month in office in 2017 he approved the Dakota pipeline, which is a 1,172 mile long pipeline for which construction was blocked in 2016 by the Obama administration. Overall, this figure provides evidence that Democrats have been more likely to inhibit the transportation of gas than Republicans in recent decades.

Aside from pipeline construction, there are a host of actions taken by recent administrations that approach the energy and clean energy sectors from different angles. Perhaps most symbolically, recent administrations have estimated the social cost of carbon, which is a measure intended to help policymakers weigh the economic benefits and drawbacks of energy related policies that might increase or decrease emissions. This measure was originally introduced under the Obama administration at \$43 a per metric ton of C02. The Trump administration's estimate was \$3 - \$5 a ton, 88 - 93 percent(!) lower than the original estimate. Finally, the Biden administration's estimate was approximately \$51 a ton. These vast differences indicate that, at least for the past few administrations, there is a stark divide between the way Republicans and Democrats view the necessity of reducing emissions. These differences have also led to a notable divergence in the types of measures taken by each party. Table~\ref{taba1} provides a comprehensive list of significant energy-related actions taken by recent administrations. Actions include, but are not limited to, international emissions reduction agreements, direct investments, tax incentives, vehicle emissions standards, loans, and royalties and leases for drilling on public lands.

\input{tablea1.tex}
\vspace{-5pt}
\begin{spacing}{1}
\noindent{\footnotesize{\textit{Notes:} This table lists notable energy policy actions of recent U.S. administrations. Where relevant, concise descriptions of each action are provided. *Estimates from The White House Briefing Room: \\ https://web.archive.org/web/20250119210015/https://www.whitehouse.gov/briefing-room/}}
\end{spacing}


\section{Iowa Electronic Markets Data}
\label{secb}

\counterwithin*{equation}{section}
\renewcommand\theequation{\thesection\arabic{equation}}
\setcounter{figure}{0}
\renewcommand{\thefigure}{B\arabic{figure}}
\setcounter{table}{0}
\renewcommand{\thetable}{B\arabic{table}}

\begin{figure}[!ht]
\caption{IEM Implied Election Probabilities}\centering
\par
\includegraphics[width=\linewidth]{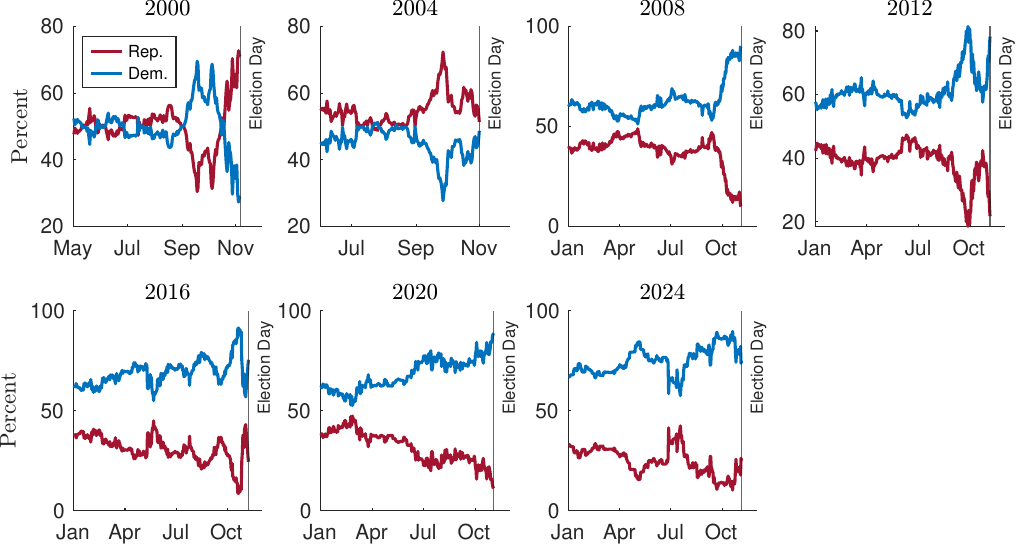} 
\begin{minipage}{1.0\textwidth}
\vspace{5pt}
\singlespace
\footnotesize \textit{Notes:} This figure plots the IEM implied presidential election probabilities for each election year in the sample.
\end{minipage}
\label{figb1}
\end{figure}

This section provides supplementary information on the data taken from Iowa Electronic Markets (IEM). IEM is operated by faculty at the University of Iowa Henry B. Tippie College of Business for research purposes. Throughout, I use prices from their winner take all U.S. presidential election markets. For each election traders may buy assets titled DEMYY\_WTA or REPYY\_WTA, where YY represents the year of the election. For example, a DEM24\_WTA contract  pays \$1 if the Democratic party nominee receives the majority of the popular vote cast in 2024, and \$0 otherwise. Thus, through the ``wisdom of the masses,'' the price of each contract on any given day is a forecast of the probability that an event will happen. To construct the probabilities that each party nominee wins the election I take the last price for the two party contracts from each day in the sample, and normalize the sum of the two prices to one.\footnote{In practice the sum of the two prices are usually within a few hundredths of one, though the existence of 3rd party contracts and low trade volumes on certain days sometimes leads to slight deviations.} For instance, on November 1, 2000, the last prices of the DEM00\_WTA and REP00\_WTA contracts were \$0.348 and \$0.668, respectively. After normalization the implied probabilities for the Democratic and Republican candidates were 34.3 and 65.7 percent, respectively. Data start May 1st, 2000 for the 2000 election, June 1st, 2004 for the 2004 election, and by at least January 1st of the election year for the remaining elections in the sample.\footnote{Throughout I only consider data starting in January of the election year.} Figure~\ref{figb1} plots the IEM implied probabilities for each election period in the sample. Vertical bars in each subplot indicate the day of the election.

\begin{figure}[!ht]
\caption{IEM Trade Volumes}\centering
\par
\includegraphics[width=\linewidth]{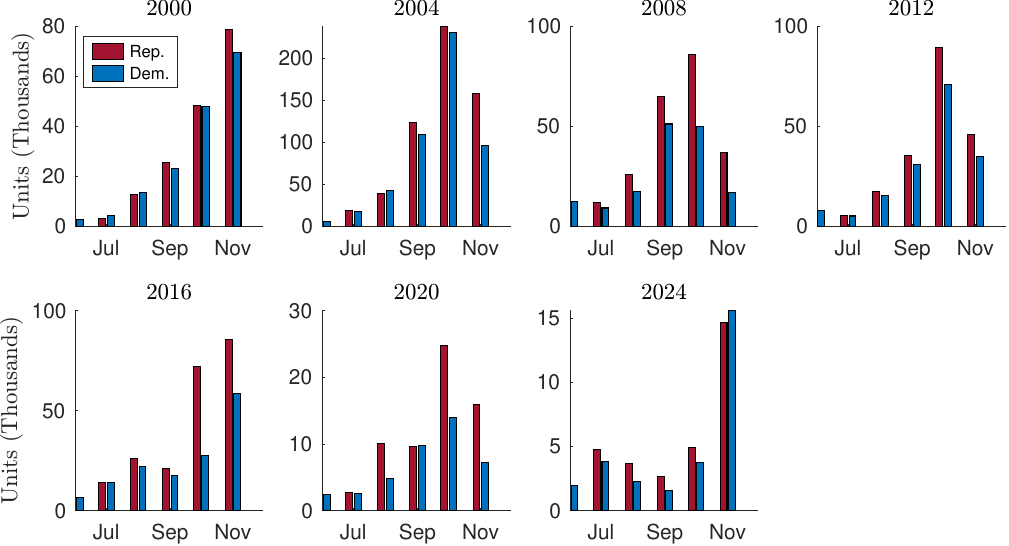} 
\begin{minipage}{1.0\textwidth}
\vspace{5pt}
\singlespace
\footnotesize \textit{Notes:} This figure plots the total IEM trade volumes for the Republican and Democratic parties in each month of the given election year.
\end{minipage}
\label{figb2}
\end{figure}

IEM also provides the number of units traded for each contract type on each day. Figure~\ref{figb2} plots the monthly trade volumes for both types of contracts. Generally, October and November have the highest trade volumes, which is unsurprising given their proximity to election day and important events like debates.

\section{Election Shocks and Events}
\label{secc}

\counterwithin*{equation}{section}
\renewcommand\theequation{\thesection\arabic{equation}}
\setcounter{figure}{0}
\renewcommand{\thefigure}{C\arabic{figure}}
\setcounter{table}{0}
\renewcommand{\thetable}{C\arabic{table}}

This section provides supplementary details on the estimation and properties of the election shocks and narrative election shocks. The estimates from equation~(\ref{eq1}) are provided in Table~\ref{tabc1}. Columns indicate variables, while rows indicate the horizon of the regressor. For example, the first row and column labeled $t-2$ gives the coefficient of $\pi^R_{t-2}$. Newey-West HAC standard errors are reported in parentheses below each estimate. The coefficients of the first two lags of $\pi^R_t$ are approximately 0.79, and 0.25 reflecting the high degree of serial correlation in the original probabilities. Contemporaneous estimates of $\Delta \text{emp}_t$ and $\Delta \text{emp}_t \times \text{Pres}^R_t$ are approximately -4.73 and 4.85, implying that when the incumbent administration is Democrat an employment report release with 0.1 percent growth leads to a 0.473pp decrease in the Republican candidate's odds. In summary, intuitively, contemporaneous news about positive employment growth benefits the incumbent party's chances at winning reelection. However, this estimate, along with the other contemporaneous financial and economic news indicators are statistically insignificant, suggesting limited overall importance for these factors impacting election probabilities. Given these estimates, in unreported results I simply take the 1-day change in $\pi^R_{t}$ as an alternative and the results of the paper remain unchanged.

\input{tablec1.tex}

\begin{figure}[!ht]
\caption{Election Shock Panels}\centering
\par
\includegraphics[width=\linewidth]{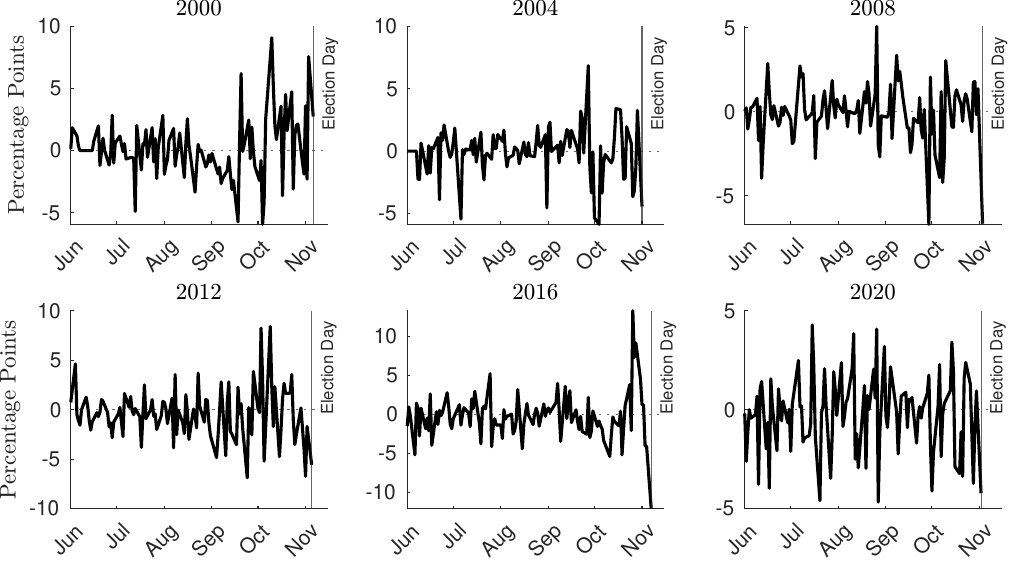} 
\begin{minipage}{\textwidth}
\vspace{5pt}
\singlespace
\footnotesize \textit{Notes:} This figure plots $\text{Shock}_t$ separately for each election period in the sample, excluding election day for ease of viewing. 
\end{minipage}
\label{figc1}
\end{figure}

Figure~\ref{figc1} plots non-election day values of $\text{Shock}_t = \hat{\varepsilon}_t$ for each election year in the sample. There is substantial heterogeneity in the variance of the shocks in each election. For example, the variance of shocks in the 2016 election is more than double the variance of shocks in the 2020 election. Larger variance can be attributed to more dramatic election years, i.e., years that contained more notable exogenous events that shifted the election probabilities. In 2020 there are no notable such events, while in 2016 there are, for example, some large shocks associated with the Comey announcement on October 28.

As discussed in Section \ref{sec42}, another approach to constructing the election shock series is to use only changes near prominent election events. Table~\ref{tabc2} lists the events considered in this paper, and provides a description of each event where relevant. For each election I include each presidential and vice presidential debate, as well as any scandals that are associated with changes in election probabilities through the altering of public opinion on one of the candidates. Many of these events were already discussed in Section \ref{sec33}. Some other recent notable events are the release of the Access Hollywood tape in the 2016 election, and the White House lockdown on May 29, 2020 amidst protests in response to George Floyd's death in an encounter with a police officer.

\input{tablec2.tex}
\vspace{-5pt}
\begin{spacing}{1}
\noindent{\footnotesize{\textit{Notes:} This table provides the dates, and where relevant, the descriptions of the narrative election events used to construct $\text{Shock}^N_t$.}}
\end{spacing}
\vspace{15pt}

\begin{figure}[H]
\caption{Narrative Shock Panels}\centering
\par
\includegraphics[width=\linewidth]{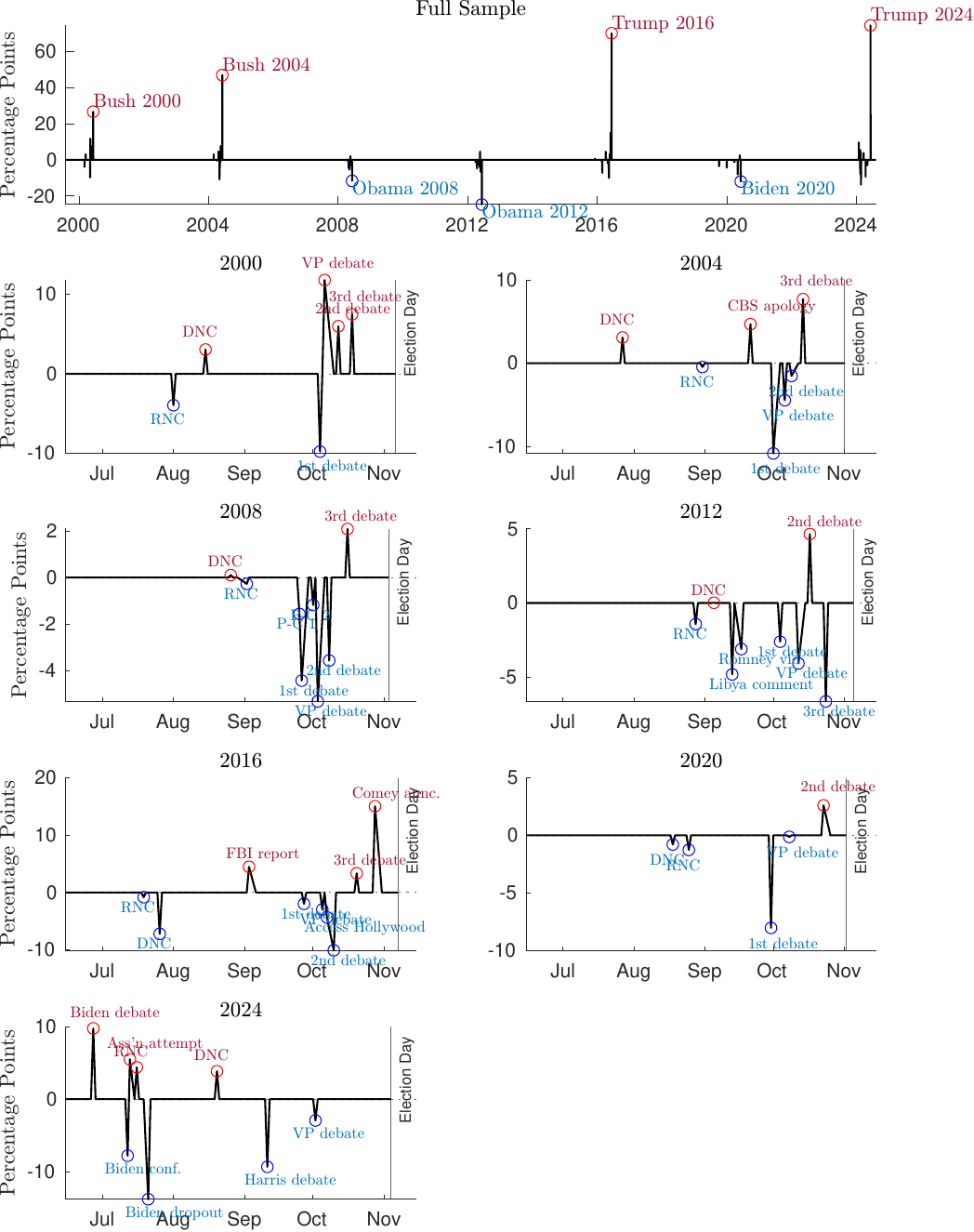} 
\begin{minipage}{1.0\textwidth}
\vspace{5pt}
\singlespace
\footnotesize \textit{Notes:} The top panel of this figure plots the narrative election shocks for the full sample. Each additional panel plots the shocks separately for each election period. Positive values reflect exogenous increases in the Republican candidate's probability of winning an election. Red (Blue) circles indicate shocks in favor of Republicans (Democrats). Text is provided to briefly describes each event associated with a shock.
\end{minipage}
\label{figc2}
\end{figure}

With these narrative events in hand, the top panel of Figure~\ref{figc2} plots $\text{Shock}^N_t$, calculated using equation~(\ref{eq5}), for the full sample. As in the original election shock series, the largest shocks are associated with election days, however many important contributions come from non-election days as well. Indeed, in the 2000, 2008, 2012, and 2020 elections the sum of the absolute value of non-election day shocks was larger than that associated with the election. Detailed panels of each election cycle are also provided, starting in July 1st for ease of viewing. In each case, text is provided indicating what event took place at each date, with red (blue) text indicating a shock in favor of the Republican (Democratic) candidate. For most election periods the first notable events are the RNC and DNC. Interestingly, a party's convention is not necessarily associated with a boost in its odds despite their association with bumps in the polls. In fact, among the Republican conventions, all but the 2024 convention are associated with a decrease in the Republican candidate's probability of winning. In short, it is likely that election betting markets take into account typical public reactions to national conventions, which may not always be associated with any lasting effect. The largest convention shock in the sample is that of the 2016 DNC, which is associated with a 7.2pp increase in favor of Clinton. 

Overall, the sign of each shock is intuitive given the narrative information at each date. For example, the Palin-Couric interviews in 2008, Romney's Libya comment and leaked audio scandals in 2012, and the Access Hollywood tape in 2016 are associated with shocks in favor of Democratic candidates. Similarly to Figure~\ref{figc1}, the Comey announcement is associated with the largest shock of 15.1pp in favor of Trump. A lot of shifts happen around presidential and VP debates, and the shocks tend to align with who was considered the ``winner'' at the time. One exception is the 1st debate in 2000, where Gore was criticized for a series of sighs. One likely explanation is that Al Gore’s performance was initially considered favorable due to this perceived superior
policy knowledge, while reactions to his negative body language took a few days to surface.
Due to the mere single day gap between the 1st debate and the VP debate, the positive values associated with the latter likely reflect changes in the reaction of voters and analysts
to the former.

\section{Sectoral Effects}
\label{secd}

\counterwithin*{equation}{section}
\renewcommand\theequation{\thesection\arabic{equation}}
\setcounter{figure}{0}
\renewcommand{\thefigure}{D\arabic{figure}}
\setcounter{table}{0}
\renewcommand{\thetable}{D\arabic{table}}

This section provides additional results to supplement the discussion in Section \ref{sec4}.

\subsection{Stock Market Effects}
\label{secd1}

\begin{figure}[!ht]
\caption{Response of $\pi^R_t$ to a 10pp Shock}\centering
\par
\includegraphics[width=0.7\linewidth]{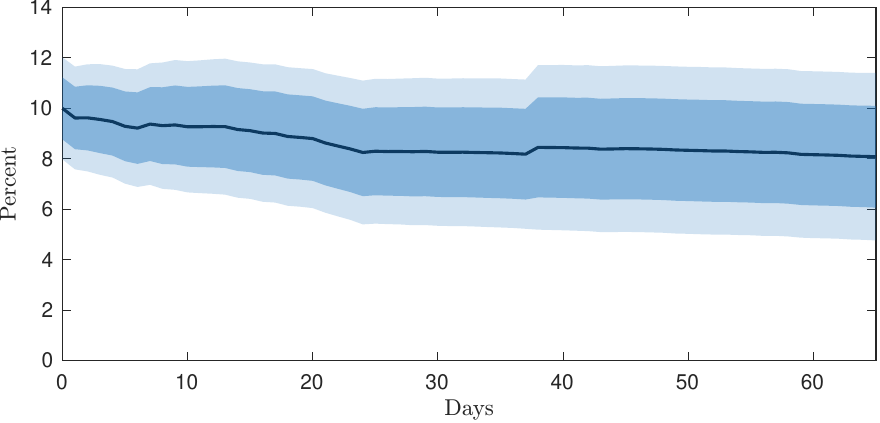} 
\begin{minipage}{1.0\textwidth}
\vspace{5pt}
\singlespace
\footnotesize \textit{Notes:} This figure plots the estimated response of $\pi^R_t$ to an election shock, normalized to increase $\pi^R_t$ by 10pp on impact. Solid black lines give point estimates, while the dark and light shaded areas give the 68 and 90 percent confidence intervals, respectively.
\end{minipage}
\label{figd1}
\end{figure}

Figure~\ref{figd1} plots the dynamic response of $\pi^R_t$, found by estimating the equation:
\begin{align}
    \pi^R_{t+h} = \beta^i_h + \gamma^i_h \text{Shock}_t + \eta^i_{t+h},
\end{align}
weighting by trade volumes as in Section \ref{sec41}. Shocks are normalized to increase $\pi^R_t$ by 10pp. Overall, the initial effects of the shock are persistent with the estimated coefficients at the furthest horizon still remaining statistically greater than zero and point estimates only dropping by roughly percent . These dynamics are partly mechanical: $\pi^R_t$ persistently goes to zero or one in the period just after an election. Thus, a shock at time $t$ will trivially be associated with persistent effects at horizon $h$ when $t+h$ takes places after election day. In unreported results I re-estimate the dynamic response of $\pi^R_t$ without including horizons that take place after the election and the response remains persistent, albeit with more volatile point estimates and larger confidence bands.

\begin{figure}[!ht]
\caption{Response to a 10pp Shock, Crude}\centering
\par
\includegraphics[width=\linewidth]{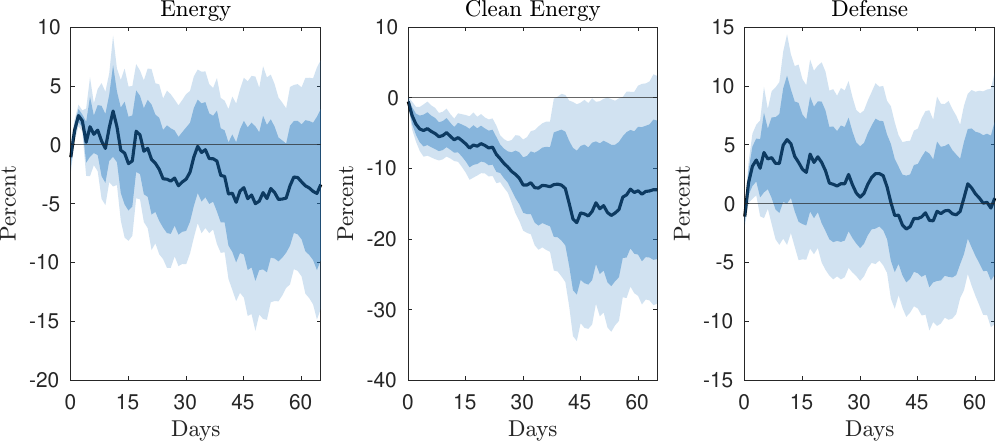} 
\begin{minipage}{1.0\textwidth}
\vspace{5pt}
\singlespace
\footnotesize \textit{Notes:} This figure plots the estimated impulse responses of sectoral stock prices to a Republican election victory, i.e., a variable equal to 1 (-1) on election days that the Republican (Democratic) candidate wins. Solid black lines give point estimates, while the dark and light shaded areas give the 68 and 90 percent confidence intervals, respectively.
\end{minipage}
\label{figd2}
\end{figure}

An alternative approach to the one explored in this paper is to estimate to election events, while ignoring the election probabilities. In particular one may estimate the equation:
\begin{equation}
    \Delta y^i_{t+h} = \beta^i_h + \upsilon^i_h \text{Elec}_t + \delta^i_h \Delta y^{i, 1m}_{t-1} + \eta^i_{t+h},
\end{equation}
where $\text{Elec}_t$ equals -1 (1) on dates that a Democrat (Republican) wins a presidential election and 0. As in $\text{Shock}_t$, this variable treats Democratic and Republican victories symmetrically, but ignores the probabilities assigned to each event. In the main text I refer to this methodology as the ``crude'' approach. The advantage of this approach is its simplicity. However there are some substantial drawbacks. First, the approach misses any innovations in election probabilities leading up to election day, reducing the sample of non-zero innovations to seven. Second, it weights each election day outcome the same regardless of the probability of the event. For instance, Obama's expected victory in 2008 is treated symmetrically to Trump's unexpected victory in 2016. If election probabilities are already factored into the price of sectoral stock prices, as narrative evidence provided in Figure~\ref{fig3} suggests, this methodology fails to account for important anticipatory effects. Figure~\ref{figd2} plots the response of each sectoral stock price to a Republican election victory. Each price responds in the expected direction contemporaneously, however in the medium-term the point estimates for Energy and Defense are near zero, or even below zero, albeit to a nonsignificant degree. The response of clean energy has a comparable shape to that of the main results displayed in Figure~\ref{fig6}. Overall, the failure to find persistent effects in Energy and Defense indicates that it is important to consider the information provided by election probabilities throughout an election cycle.

\begin{figure}[!ht]
\caption{Response to a 10pp Shock, One-Step}\centering
\par
\includegraphics[width=\linewidth]{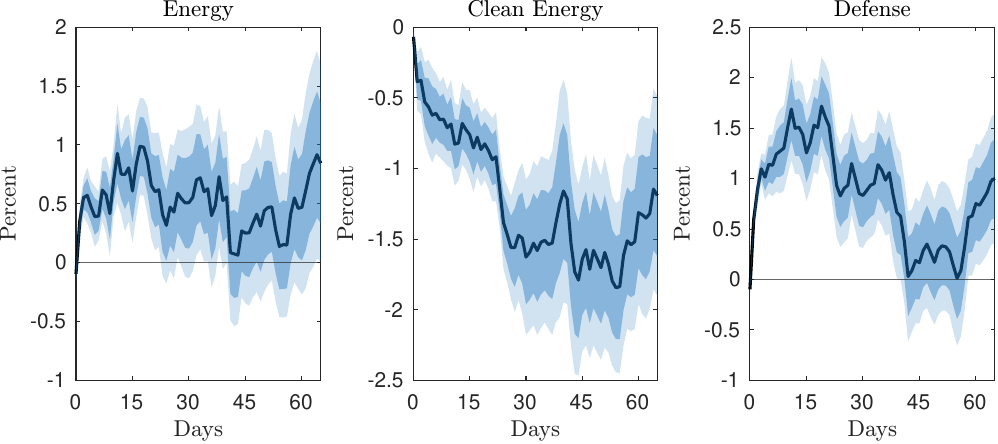} 
\begin{minipage}{1.0\textwidth}
\vspace{5pt}
\singlespace
\footnotesize \textit{Notes:} This figure plots the estimated impulse responses of sectoral stock prices to a 10pp increase in $\pi^R_t$ under the one-step approach discussed in Section \ref{secd1}. Solid black lines give point estimates, while the dark and light shaded areas give the 68 and 90 percent confidence intervals, respectively.
\end{minipage}
\label{figd3}
\end{figure}

As a robustness check to the main results 4.1, I also take a one-step approach to estimating the effects of an exogenous increase $\pi^R_t$ without estimating $\text{Shock}_t$. To do so I estimate the equation:
\begin{align}
    \Delta y^i_{t+h} & = \alpha^i_0 + \sum_{s=0}^{5}\alpha^i_{1, s}\pi^{\text{R}}_{t-s} + \sum_{s=0}^{5}\alpha^i_{2, s}\textbf{X}_{t-s} + \alpha^i_3 \text{Pres}^{\text{R}}_{t} \nonumber \\ 
    & + \sum_{s=0}^{5}\alpha^{i}_{4, s}{'}\left( \textbf{X}_{t-s} \times \text{Pres}^{\text{R}}_{t}\right) + \delta^i_h \Delta y^{i, 1m}_{t-1} + \varepsilon^i_{t+h},
    \label{eqd3}
\end{align}
where each variable is defined the same way as Sections \ref{sec3} and \ref{sec4}, and the coefficient of interest is $\alpha^i_{1, 0}$, i.e., the contemporaneous effect of a change in $\pi^R_t$. As opposed to the baseline specification, this approach does not suffer from a potential generated regressor problem. The identifying assumption is that after controlling for the other included regressors, changes to $\pi^R_t$ are exogenous to $\Delta y^i_{t+h}$. I estimate equation~(\ref{eqd3}) using weighted OLS. Figure~\ref{figd3} plots the response of each sectoral stock price to a 10pp increase to $\pi^R_t$. The responses largely mimic those of the main results plotted in Figure~\ref{fig6}.

\begin{figure}[!ht]
\caption{Response to a 10pp Shock, Removed 2008 and 2020}\centering
\par
\includegraphics[width=\linewidth]{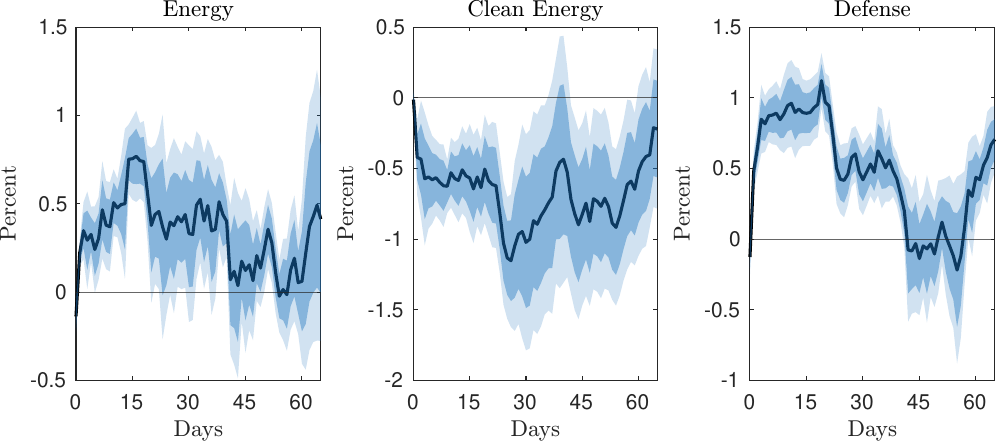} 
\begin{minipage}{1.0\textwidth}
\vspace{5pt}
\singlespace
\footnotesize \textit{Notes:} This figure plots the estimated impulse responses of sectoral stock prices to an election shock, normalized to increase $\pi^R_t$ by 10pp. 2008 and 2020 values are removed to exclude the Great Recession and Covid-19 periods. Solid black lines give point estimates, while the dark and light shaded areas give the 68 and 90 percent confidence intervals, respectively.
\end{minipage}
\label{figd4}
\end{figure}

One identification challenge present throughout the paper is the limited sample of only seven election periods. If large swings in sectoral stock prices were caused by other major events that took place in the same year as an election, the estimated responses to election shocks may be confounded with other important influences. The two most feasible such events are the 2007-2008 financial crisis and the 2020 COVID-19 pandemic. To illustrate, consider a large negative election shock in 2008, perhaps Obama's victory in the election. If energy prices sharply and unexpectedly decreased after the election due to the reasons associated with the financial crisis, we might attribute this decrease to Obama's victory instead. To investigate this concern, I re-estimate equation~(\ref{eq4}) while dropping election shocks from 2008 and 2020, and plot the results in Figure~\ref{figd4}. Interestingly, the point estimates remain similar to that of the baseline results, though the confidence bands are generally tighter likely due to the removal of particularly volatile data points from the 2008 and 2020 election. 

\subsection{Labor Market Effects}
\label{secd2}

In the main text I study the response of jobs in the industries that I label oil drilling \& extraction, mining \& quarrying, clean energy generation, aerospace manufacturing, and ship \& tank manufacturing. Table~\ref{tabd1} provides the North American Industry Classification System (NAICS) codes representing each of these categories in the data.

\input{tabled1.tex}
\vspace{-20pt}
\noindent{\footnotesize{\textit{Notes:} This table lists the NAICS codes that comprise each industry used in the paper.}}

\begin{figure}[!ht]
\caption{Response to a 10pp Shock, Defense}\centering
\par
\includegraphics[width=\linewidth]{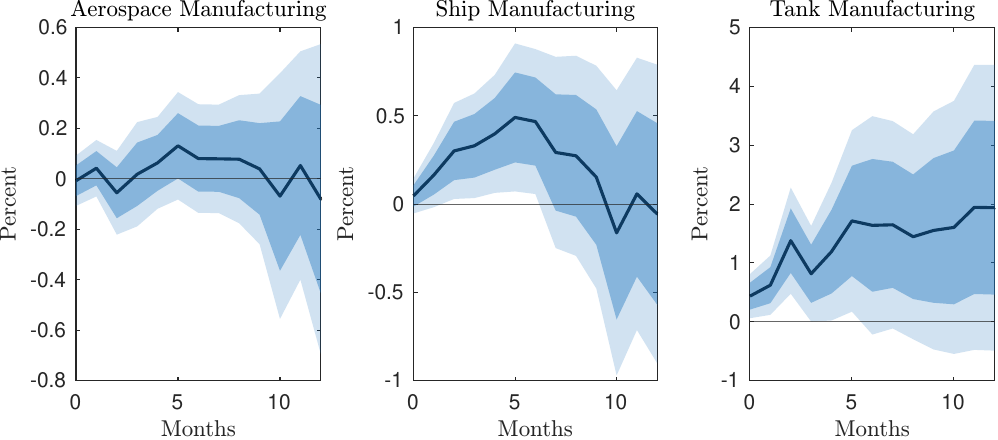} 
\begin{minipage}{1.0\textwidth}
\vspace{5pt}
\singlespace
\footnotesize \textit{Notes:} This figure plots the estimated defense manufacturing employment responses to a 10pp election shock. Solid black lines give point estimates, while the dark and light shaded areas give the 68 and 90 percent confidence intervals, respectively.
\end{minipage}
\label{figd5}
\end{figure}

For brevity in the main text, I combine ship and tank manufacturing, since the two industries together still make up only roughly a quarter of defense manufacturing jobs in the sample. Figure~\ref{figd5} plots the dynamic response of ship and tank manufacturing employment separately, re-including aerospace manufacturing for ease of viewing. In the short-term positive election shocks, i.e., "Republican shocks", increase both ship and tank manufacturing. However these effects only appear to be persistent for tank manufacturing jobs.

\end{document}

%% file: tablea1.tex
\begin{longtable}{lllp{7cm}}
\caption{Energy Policy of U.S. Administrations} \\
    \textbf{Date} & \textbf{Admin} & \textbf{Action} & \textbf{Description} \\
    \hline
    12/11/97 & Clinton & Kyoto Protocol Signed & An international agreement to reduce C02 emissions. \\
    3/28/01 & Bush  & Kyoto Protocol Abandoned &  \\
    8/5/05 & Bush  & Energy Policy Act of 2005 & Key provisions were investment in cleaner coal plants, tax incentives to expand pipelines, requirement for gas to contain ethanol, and tax credits for solar energy. \\
    3/8/08 & Bush  & CA Emissions Waiver & Denied the California Emissions Waiver, an an attempt to set stricter vehicle emissions standards for new cars in California than required by Federal law. \\
    2/17/09 & Obama & ARRA 2009 & Contributed an estimated \$90 billion* in clean energy investment and contained some smaller incentives for carbon capture. \\
    7/8/09 & Obama & CA Emissions Waiver & Approved \\
    12/18/09 & Obama & Copenhagen Accord & Agreed to an international agreement to reduce C02 emissions. \\
    2/16/10 & Obama & Vogtle Nuclear Plant Loan & \$8.3 billion loan guarantee to Plant Vogtle for the construction of two new reactors. \\
    5/7/10 & Obama & 2012-2016 Light VES & Emissions standards for light-duty vehicels, which required 2012-2016 models to meet final average emissions level of 35.5 mpg. \\
    11/15/11 & Obama & 2014-2018 Heavy VES & Emissions standards for heavy-duty vehicles, which required 2014-2018 models to to reduce C02 missions by an estimated 270 million metric tons. \\
    10/15/12 & Obama & 2017-2025 Light VES & Restricts 2017-2025 models to meet final average emissions level of 54.5 mpg by 2025. \\
    8/3/15 & Obama & Clean Power Plan & Required each state to reduce carbon emissions within its borders by a target amount. \\
    11/6/15 & Obama & Keystone XL Rejected & Rejected a proposed 4th Keystone pipeline for oil transportation. \\
    12/18/15 & Obama & Oil Export Ban Repealed & Removed 1975-2015 legislation that prevented most crude oil exports from the U.S. to other countries. \\
    9/3/16 & Obama & PCA Entered & An international treaty to reduce C02 emissions. \\
    10/25/16 & Obama & 2021-2027 Heavy VES & Restricts 2021-2027 models to to lower C02 emissions by an estimated 1.1 billion metric tons. \\
    12/4/16 & Obama & Dakota Pipeline Blocked & Blocked oil pipeline due to it crossing the Sioux tribe's land. \\
    2/17/17 & Trump & Dakota Pipeline Allowed & Allowed pipeline. On May 14, 2017, the first oil was sent through the pipeline. \\
    3/24/17 & Trump & Keystone XL Revived & Presidential permit to allow bulding of the pipeline. \\
    6/1/17 & Trump & PCA Withdrawal &  \\
    8/24/18 & Trump & Amend Light VES 2021-2026 & Blocked tightening of C02 emissions standards for 2021-2016 models \\
    9/12/18 & Trump & EPA Budget Cut & 31\% cut to EPA's budget. \\
    9/28/18 & Trump & NEICA 2017 & Cut regulatory costs to expand nuclear research progress. \\
    6/19/19 & Trump & Clean Power Plan Repealed & Replaced by the ACE rule, which had less stringent requirements on carbon emissions. \\
    9/19/19 & Trump & CA Waiver Revocation &  \\
    1/20/21 & Biden & Keystone XL Revoked & Revoked the permit granted by Trump, and in June of 2021 the project was abandoned. \\
    1/20/21 & Biden & PCA Re-Entered &  \\
    1/27/21 & Biden & Pause Oil/Gas Leases & Executive Order to halt leases for drilling on public lands. \\
    11/15/21 & Biden & IIJA  & Includes provisions to electrify school buses, improve transportation alternatives like biking and walking, and \$65 billion* in clean energy transmission. \\
    12/30/21 & Biden & 2023-2026 Light VES & Average emissions requirement of 55mpg by 2026. \\
    5/12/22 & Biden & Canceled Leases & Abolished three drilling leases in the Gulf of Mexico and Alaska. \\
    8/16/22 & Biden & IRA of 2022 & Estimated to include \$265 billion* in clean energy investments and tax incentives. \\
    4/18/24 & Biden & 2027-2032 Light VES & Tightening of standards (varying by vehicle type) that are estimated to reduce emissions by 7.2 billion metric tons. \\
    4/12/24 & Biden & Revised BLM Regulations & The Bureau of Land Management tightened its oil and gas leasing regulations by increasing royalties from 12.5 to 16.67 percent, increasing minimum bids for land, base rental rates, and adding a fee for expressing interest. \\
    4/22/24 & Biden & 2027-2032 Heavy VES & Tightening of standards that are estimated to reduce emissions by 1 billion metric tons. \\
    1/6/25 & Biden & Offshore Oil/Gas Ban & Ban on drilling of 625 million acres of ocean. \\
    1/20/25 & Trump & PCA Withdrawal &  \\
    1/20/25 & Trump & Other Executive Orders & Several actions related to energy including EPA reviews of regulations related to C02 emissions, lifting of restrictions on oil and gas developement, paused leasing for wind projects, and a pause on IRA and IIJA funds for eletric vehicle goals, and removal of environmental justice considerations. \\
    \hline
    \label{taba1}
\end{longtable}

%% file: tablec1.tex
\begin{table}[!ht]
\begin{threeparttable}
  \centering
  \caption{Election Shock Regression}
    \begin{tabular}{lllllll}
    Variable\textbackslash{}Lag & $t$ & $t-1$ & $t-2$ & $t-3$ & $t-4$ & $t-5$ \\
    \hline
    $\pi^R$ &       & 0.789*** & 0.249* & -0.049 & 0.044 & -0.029 \\
          &       & (0.167) & (0.136) & (0.094) & (0.100) & (0.062) \\

    $\text{Pres}^R$ & -0.002 &       &       &       &       &  \\
          & (0.003) &       &       &       &       &  \\

    $\Delta i$ & 0.055 & 0.03  & 0.02  & 0.012 & 0.061 & -0.017 \\
          & (0.043) & (0.031) & (0.036) & (0.024) & (0.037) & (0.022) \\

    $\Delta \text{sp500}$ & 0.013 & 0.417 & -0.174 & -0.292 & -0.312** & 0.103 \\
          & (0.187) & (0.297) & (0.106) & (0.281) & (0.132) & (0.177) \\

    $\Delta \text{emp}$ & -4.731 & 0.795 & 12.931 & 1.729 & 0.233 & 0.073 \\
          & (4.142) & (5.082) & (12.471) & (3.185) & (3.230) & (3.350) \\

    $\Delta \text{cpi}$ & -2.381 & 0.583 & 1.101 & 0.132 & -2.436** & 1.162 \\
          & (1.828) & (1.556) & (1.533) & (1.519) & (1.195) & (1.363) \\

    $\Delta \text{ind}$ & -0.183 & 0.509 & 0.542 & -0.195 & 0.002 & -0.101 \\
          & (0.787) & (0.622) & (0.734) & (0.790) & (0.559) & (0.676) \\

    $\Delta i \times \text{Pres}^R$ & -0.035 & 0.003 & -0.024 & -0.017 & -0.041 & 0.023 \\
          & (0.045) & (0.034) & (0.039) & (0.027) & (0.041) & (0.027) \\

    $\Delta \text{sp500} \times \text{Pres}^R$ & -0.009 & -0.441 & 0.243** & 0.384 & 0.332** & -0.062 \\
          & (0.194) & (0.291) & (0.114) & (0.293) & (0.158) & (0.209) \\

    $\Delta \text{emp} \times \text{Pres}^R$ & 4.848 & -0.722 & -12.857 & -1.831 & -0.237 & -0.571 \\
          & (4.133) & (5.093) & (12.459) & (3.186) & (3.235) & (3.365) \\

    $\Delta \text{cpi} \times \text{Pres}^R$ & 2.027 & 1.393 & -0.75 & -0.29 & 2.378 & -2.733* \\
          & (1.892) & (2.001) & (1.763) & (1.712) & (1.535) & (1.640) \\

    $\Delta \text{ind} \times \text{Pres}^R$ & 0.752 & -0.461 & -0.484 & 0.05  & -0.031 & 0.111 \\
          & (0.888) & (0.630) & (0.737) & (0.803) & (0.565) & (0.682) \\

    \hline
    Observations & 1,259 & & & & & \\
    R-Squared & 0.885 & & & & & \\
    \hline
    \end{tabular}%
    \begin{tablenotes}[flushleft]
        \item {\footnotesize{\textit{Notes:} This table reports the coefficients from the OLS estimate of equation (1). Newey-West HAC consistent estimator with a bandwidth of $\frac{3}{4}\text{T}^\frac{1}{3}$ are reported in parentheses under each point estimate. *$p<0.10$, **$p<0.05$, ***$p<0.01$.}}
    \end{tablenotes}
    \label{tabc1}
  \end{threeparttable}
\end{table}%

%% file: tablec2.tex
\begin{longtable}{llp{9cm}}
  \caption{Election Events} \\
    \textbf{Date} & \textbf{Event} & \textbf{Description} \\
    \hline
    7/31/00 & RNC & \\
    8/14/00 & DNC & \\
    10/3/00 & 1st debate &  \\
    10/5/00 & VP debate &  \\
    10/11/00 & 2nd debate &  \\
    10/17/00 & 3rd debate &  \\
    11/7/00 & Election &  \\
    7/26/04 & DNC & \\
    8/30/04 & RNC & \\
    9/20/04 & CBS Apology & CBS issues an apology about illegitmate documents in a negative story of Bush's military service. \\
    9/30/04 & 1st debate &  \\
    10/5/04 & VP debate &  \\
    10/8/04 & 2nd debate &  \\
    10/13/04 & 3rd debate &  \\
    11/2/04 & Election &  \\
    8/25/08 & DNC & \\
    9/1/08 & RNC & \\
    9/25/08 & Palin-Couric 1 & Release of an interview with McCain's running mate, Sarah Palin, who was criticized for several comments. \\
    9/26/08 & 1st debate &  \\
    9/30/08 & Palin-Couric 2 & Release of additional segment. \\
    10/2/08 & VP debate &  \\
    10/7/08 & 2nd debate &  \\
    10/15/08 & 3rd debate &  \\
    11/4/08 & Election &  \\
    8/27/12 & RNC & \\
    9/4/12 & DNC & \\
    9/12/12 & Libya/Egypt Comment & Romney criticized for response to Obama's handling of attacks on U.S. embassies in Egypt and Libya. \\
    9/17/12 & Romney Video & Leaked Romney video where he discusses "entitled" Obama voters to donors. \\
    10/3/12 & 1st debate &  \\
    10/11/12 & VP debate &  \\
    10/16/12 & 2nd debate &  \\
    10/23/12 & 3rd debate &  \\
    11/6/12 & Election &  \\
    5/3/16 & Cruz Drops out & Cruz dropped out of the Republican primary. Kasich also dropped out  a day prior, making Trump the presumptive nominee. \\
    7/18/16 & RNC & \\
    7/25/16 & DNC & \\
    9/2/16 & FBI Email Report & An email report about the investigation of Clinton's misuse of personal emails for work was released. \\
    9/26/16 & 1st debate &  \\
    10/4/16 & VP debate &  \\
    10/7/16 & Access Hollywood & A tape with audio of Trump discussing sexually assaulting
women was released. \\
    10/9/16 & 2nd debate &  \\
    10/19/16 & 3rd debate &  \\
    10/28/16 & Comey Announcement & James Comey (director of the FBI) announced that the reopening of an investigation of Clinton's email misuse. \\
    11/8/16 & Election &  \\
    3/3/20 & Super Tuesday & Biden performed well on super Tuesday, and democratic odds increased likely due to a perception of his chances against Trump. \\
    5/29/20 & George Floyd Lockdown & The White House locked down due to protests, which Trump was criticized for. \\
    8/17/20 & DNC & \\
    8/24/20 & RNC & \\
    9/29/20 & 1st debate &  \\
    10/7/20 & VP debate &  \\
    10/22/20 & 2nd debate &  \\
    11/3/20 & Election &  \\
    6/27/24 & Biden debate &  \\
    7/11/24 & Biden Conference & Biden held a news conference, his first public appearance speaking since a tumultuous debate. \\
    7/13/24 & Ass'n Attempt & An attempt on Trump's life was made at a rally in Pennsylvania. \\
    7/15/24 & RNC & \\
    7/21/24 & Biden Dropout & Joe Biden dropped out of the race and endorsed Kamala Harris. \\
    8/19/24 & DNC & \\ 
    9/10/24 & Harris debate &  \\
    10/1/24 & VP debate &  \\
    11/5/24 & Election &  \\
    \hline
    \label{tabc2}
\end{longtable}%

%% file: tabled1.tex
\begin{table}[H]
  \centering
  \caption{Industry Classifications}
    \begin{tabular}{ll}
    \textbf{Industry} & \textbf{NAICS Codes} \\
    \hline
    Oil Drilling \& Extraction & 211, 213111, 213112 \\
    Mining \& Quarrying & 212, 213113, 213114, 213115 \\
    Clean Energy Generation & 221111, 221112, 221113, 221114, 221115, 221116 \\
    Aerospace Manufacturing & 3364 \\
    Ship Manufacturing & 336992 \\
    Tank Manufacturing & 3366 \\
    \hline
    \end{tabular}%
    \label{tabd1}
\end{table}%